\documentclass[useAMS,usenatbib]{mn2e}
\usepackage{graphicx}
\usepackage{float}
\usepackage[dvips]{color}
\usepackage[usenames,dvipsnames]{xcolor}
\usepackage[normalem]{ulem}
\usepackage{subfigure}
\usepackage{lastpage}

\title[Spectroscopic survey of the thick disk-halo interface]{A 10,000 star spectroscopic survey of the thick disk-halo interface : Phase-space sub-structure in the thick disk} 
\author[Jayaraman, Gilmore, Wyse, Norris, Belokurov]{Apoorva Jayaraman$^{1}$\thanks{E-mail:aj@ast.cam.ac.uk},
  Gerard Gilmore$^{1}$, Rosemary F.G. Wyse$^{2}$, John E. Norris$^{3}$,
  \newauthor Vasily Belokurov$^{1}$ \\
$^{1}$Institute of Astronomy, Madingley Rd, Cambridge CB3 0HA, UK\\ 
$^{2}$Johns Hopkins University, Department of Physics and Astronomy, 3900 North Charles Street, Baltimore, MD 21218, USA\\
$^{3}$Research School of Astronomy and Astrophysics, Australian National University, Mount Stromlo Observatory, Cotter Road, \\Weston, ACT 2611, Australia
} 
\begin{document} 
 
\date{Accepted 2013 February 4.  Received 2013 February 4; in original form 2012 October 6} 
 
 \pagerange{\pageref{firstpage}--\pageref{lastpage}} \pubyear{1066} 
 
\maketitle 
 
\label{firstpage} 
 
\begin{abstract} 

We analyse a 10,000 star spectroscopic survey, focused on Galactic
thick disk stars typically 2-5 kpc from the Sun, carried out using the
AAOmega Spectrograph on the AAT.  We develop methods for
completeness-correction of the survey based on SDSS photometry, and we
derive star distances using an improved isochrone-fitting method with
accuracies better than 10\%. We determine the large-scale kinematic
($V_{\phi}$=172 km/s and $\sigma(\phi,r,z)$ = (49,51,40) km/s.), and
abundance properties of the thick disk, showing these representative
values are a fair description within about 3 kpc of the sun, and in
the range 1-3 kpc from the Galactic Plane. We identify a
  substantial overdensity in lines of sight towards the inner Galaxy,
  with metallicity [Fe/H]$\sim -1$ dex, and higher line of sight
  velocities than the thick disk, localised along the direction
$(\ell,b)=(48^{\circ},-26^{\circ})$. This overdensity appears to be
towards, but closer than, the known Hercules-Aquila halo overdensity
and may be related to the Humphreys-Larsen inner galaxy thick disk
asymmetry.

\end{abstract} 
 
 
\section{Introduction} 
 
The advent of large-area photometric and spectroscopic surveys in the
recent decade has made observational studies of the Galactic stellar
populations extremely vibrant. Rapid progress is being made in
obtaining independent descriptions of the chemical, kinematic and
spatial distributions, especially of the thick disk and halo, although
systematic challenges remain associated particularly with the
definitions of stellar populations. Photometric definitions, which led
to discovery of the thick disk, are reviewed by, for example,
\cite{GWK} and the history of such studies up to the present is
described by \cite{Yosh}. Much recent focus has been on chemical
abundance definitions. Those studies based on either small numbers of
high quality spectra or large numbers of lower resolution spectra,
have become a major industry. They have established that the thick
disk is a discrete stellar population, at least in the solar
neighbourhood (but {\sl cf.} \citet{Bovy} for an orthogonal view), but
raised many open questions and inconsistent results on gradients,
origins, and so on (\citet{KF1,KF2,KF3}, \citet{KF4}, \citet{BR1,BR4},
\citet{BR2}, \citet{BR3,BR5}, \citet{TB1,TB2}, \citet{TB2011,TB3},
\citet{Lee1}, \citet{KS1}, \citet{Cheng1,Cheng2}, \citet{LV12},
\citet{Greg11}). As just two more of these very many recent examples,
\citet{b9} determined the exponential scale height and scale length of
the thick disk from Sloan Digital Sky Survey (SDSS) photometric data
to be 0.9 kpc and 3.6 kpc respectively. \citet{b4} re-visited the
kinematics of the thick disk and halo using SDSS Data Release 7 (DR7)
and particularly investigated the overlaps between known stellar
populations to conclude that the metal weak thick disk is a
kinematically and chemically distinct component between the halo and
thick disk. The main challenge for these kinds of studies has been the
intrinsic overlap of population distribution functions, and the
associated complexities in how to deconvolve them correctly. Improving
constraints then requires a moderately high surface density
spectroscopic survey, with accurate photometric and proper motion
data, sampling a sufficient range of Galactic coordinates to support
statistical separation of the populations. Much is being learnt, but
much remains to be earned, especially from spectroscopy outside the
Solar neighbourhood (for example, \citet{Kordo})

We present here analysis of a 10,000-star spectroscopic survey, with
Anglo-Australian Observatory AAOmega spectroscopy complementing SDSS
photometry, that has the potential for single-population and
correlated population-overlap studies of the kinematics, chemistry and
spatial distribution of the thick disk. This survey uses F/G stars to
probe significant distances from the Sun; the chemical abundances of
these stars provide information on the enrichment history during early
stages of galaxy formation. The radial velocities, along with chemical
abundance and spatial distribution, allow us to differentiate between
populations. The main scientific objective of this survey was to
determine the unbiased distribution functions over kinematics,
metallicity and spatial distributions of the thick disk and halo, with
emphasis on the region of overlap between these two populations.

In section 2 of this paper we present the data, discuss the selection
function and compare the photometry and spectroscopy to SDSS DR7. In
section 3 we present a weighting method for completeness correction,
and thereafter explain an isochrone fitting technique used to derive
distances to the stars. In this same section, we also provide a brief
discussion of the proper motions (taken from SDSS DR7) and show that
the resulting tangential velocities have inevitable large
uncertainties. We hence show that, for this study, it is better to
exclude proper motions from the analysis and model the data with
parameters derivable from the spectroscopic measurements i.e. line of
sight velocities, metallicities and distances. In section 4 we outline
our method to implement such modelling by marginalising over unknown
parameters. We then derive and present our results for the
best-fitting kinematic and photometric model of the thick disk, and
analyse and characterise the residuals of this model when applied to
the data, searching for - and identifying - significant phase-space
substructure.

\begin{figure} 
\includegraphics[width=1.0\linewidth]{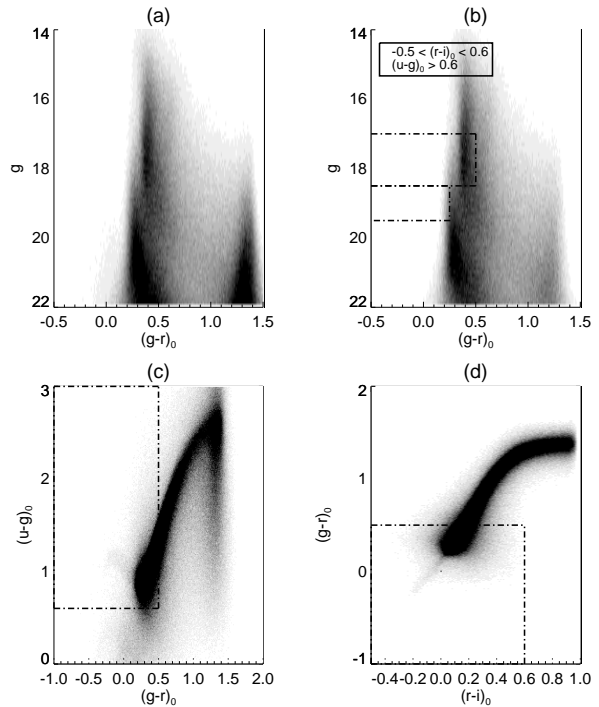} 
\caption{The WGN selection function to isolate stars from the thick disk-halo  
interface \citep{Datainprep} [a] colour-magnitude diagram of data from SDSS DR7  
covering 32 fields spanning $-1^{\circ} < \mathrm{Dec} < 1^{\circ}$ and 
$133^{\circ} <   \mathrm{RA} < 339^{\circ}$, and a broad selection function : 
$14 \le g_0 \le 22$, $-1 \le (g-r)_0 \le 1.7$, $-1 \le (r-i)_0 \le 1$, $(u-g)_0 \ge 0$. 
The main   Galactic stellar populations seen in this CMD are - the thin disk  
 near $(g-r)_0 \sim 1.3$, the thick disk (brighter) and halo (fainter)   
near $(g-r)_0 \sim 0.4$; [b] CMD of (a) with the following selection :  
$-0.5<(r-i)_0<0.6$, $(u-g)_0\ge0.6$. This significantly reduces thin disk   
contamination. The area enclosed within the dot-dashed box denotes the region 
that is selected from $g$ vs $(g-r)_0$ space to isolate stars   from the thick 
disk-halo interface; [c,d] Colour-colour diagram for the same data as in (a). 
The $(u-g)_0$ cut mainly eliminates quasars while the $(g-r)_0$ and $(r-i)_0$ 
cuts eliminate thin disk and binary stars. The dot-dashed box again indicates 
the selected region of thick  disk-halo interface stars isolated for study here. }
\label{SDSScmd} 
\end{figure}

\begin{figure} 
\includegraphics[width=1.0\linewidth]{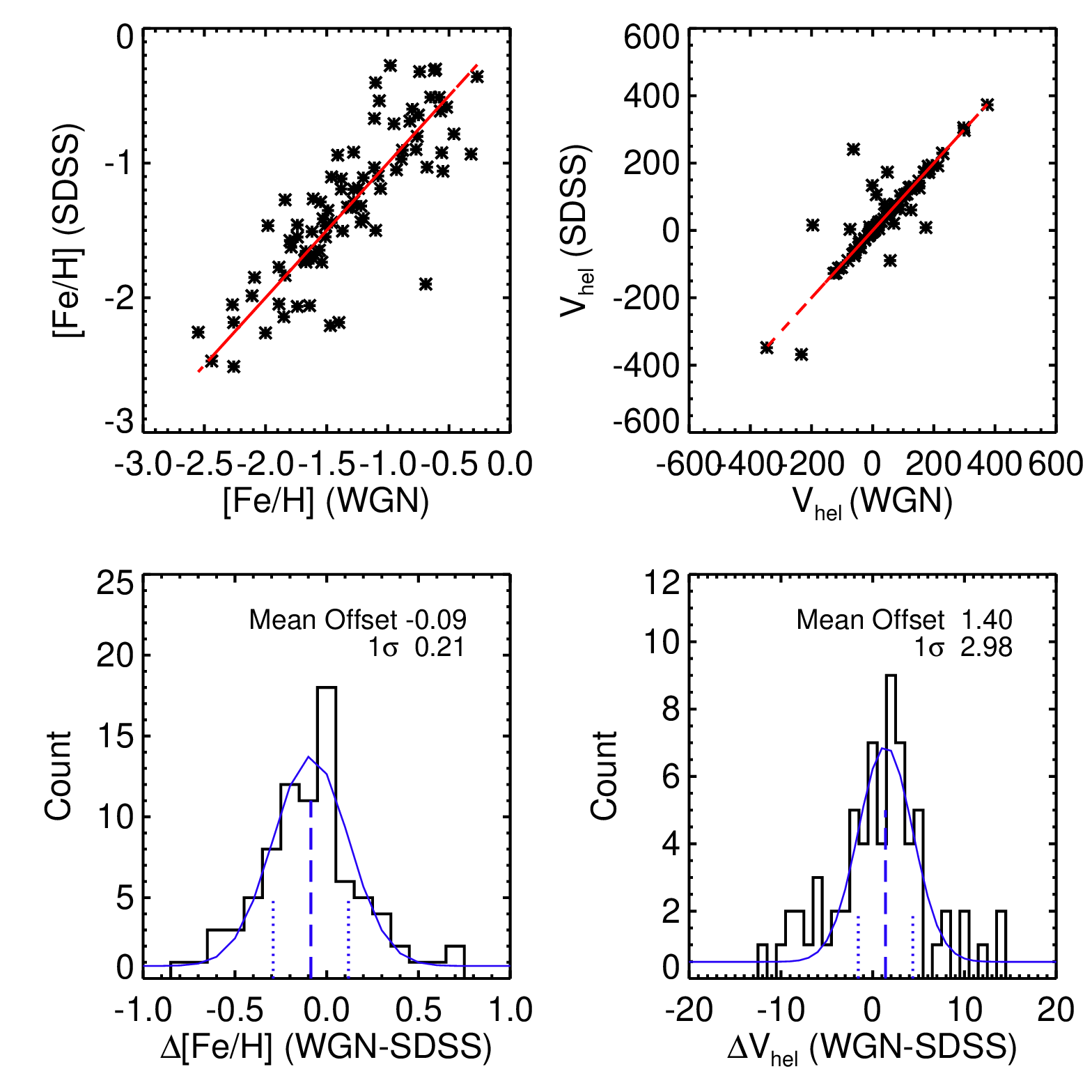} 
\caption{Comparison of derived quantities from  
Spectroscopy from WGN (AAOmega Survey) to the
  cross-matched spectroscopy from SDSS DR7. 84 matches are found based
  on the position of stars. The top 2 panels show scatter plots of
  [Fe/H] and radial velocity from WGN against the values from SDSS
  DR7, where the dashed line indicates the one-to-one correlation for
  WGN data. The bottom panels show the distribution of the difference
  between the two measurements. The black histogram denotes the
  difference between WGN and SDSS DR7, the blue smooth line shows a
  gaussian fit to this histogram. The dashed vertical line in each
  case marks the mean offset of SDSS DR7 spectroscopy from WGN, while
  the dotted vertical lines marks the $1\sigma$ limits of the
  scatter. The value of the mean offset and $1\sigma$ limits for both
  [Fe/H] and radial velocity is indicated in the respective plots.}
\label{spectrophoto_compare} 
\end{figure} 

\section[THE DATA]{THE DATA} 
\subsection{The Survey} 
A wide-field spectroscopic survey was carried out by
\citet[][(hereafter WGN)]{b101,Datainprep} using the AAOmega
Spectrograph on the Anglo Australian Telescope, fed by the 400-fibre
2dF facility, targeting lines of sight in the thick disk-halo
interface, each covering a two degree field of view. Spectroscopy was
obtained for $\sim 10,000$ stars, drawn from the Sloan Digital Sky
Survey (SDSS) photometric catalogue, along 32 lines of sight in the
region $133^{\circ} < \mathrm{RA} < 339^{\circ}; -1^{\circ} <
\mathrm{Dec} < 1^{\circ}$ i.e.  along the celestial equator. These
cover Galactic longitudes between $228^{\circ}$ and 67$^{\circ}$ at
intermediate latitudes ($26^{\circ} < |b| < 63^{\circ}$). Table
\ref{Table_fields} provides the coordinates of all the observed
lines of sight along with the number of stars analysed in each field.

\begin{table*} 
 \centering 
 \begin{minipage}{140mm} 
  \caption{Fields Covered by the WGN Data} 
  \begin{tabular}{@{}lccccc@{}} 
  \hline 
   RA,Dec     &   $l,b$      & Mean Reddening   & Number of Stars & Number of Stars  & \% Completeness of \\ 
   deg & deg & E(B-V) & analysed from WGN  & SDSS DR7 & WGN spectra wrt SDSS DR7 \\ 
 \hline 
     133 ,0  &   228, 27  &  0.033 &     288  &      2907 &      10 \\ 
     135 ,0  &   229, 28  &  0.035 &     236    &    2523 &      9 \\ 
     139 ,0  &   231,     32 & 0.030  &       254 &       1855 &      13 \\ 
     141 ,0  &   233,     33   & 0.032 &      26 &        1713 &      1 \\ 
     145 ,0  &   235,     37   & 0.064 &    268 &       1524 &      17 \\ 
     153 ,0  &   242,     43   & 0.036 &     255 &        1280 &      20 \\ 
     159 ,0  &   247,     48   & 0.065 &     304 &        1210 &      25 \\ 
     161 ,0  &   249,     49   & 0.047 &     299 &        1238 &      24 \\ 
     163 ,0  &   251,     50   & 0.046 &     130 &        1211 &      10 \\ 
     175 ,0  &   268,     58   & 0.021 &     134 &       1187 &      11 \\  
     179 ,0  &   275,     60   & 0.022 &     118 &        1255 &      9 \\ 
     185 ,0  &   286,     62   & 0.024 &     262  &       1242 &      21 \\ 
     189 ,0  &   295,     63   & 0.021 &    123  &       1376 &      9 \\ 
     197 ,0  &   312,     63   & 0.023 &     294  &       1559 &      19 \\ 
     199 ,0  &   316,     62   & 0.028 &     253  &       1546 &      16 \\ 
     203 ,0  &   324,     61   & 0.027 &     274  &       1717 &      16 \\ 
     205 ,0  &   328,     60   & 0.028 &     257  &       1790 &      14 \\ 
     213 ,0  &   341,     57   & 0.041 &     199  &       2190 &      9 \\ 
     217 ,0  &   347,     54   & 0.039 &     270  &       2428 &      11 \\ 
     223 ,0  &   354,     50   & 0.047 &     288  &      2836 &      10 \\ 
     229 ,0  &   1,     45   & 0.057 &     574  &      2669 &      15 \\ 
     235 ,0  &   6,     41     & 0.100 &   209    &    4668 &      5 \\ 
     237 ,0  &   7,     39     & 0.093 &   275    &    4859 &      5 \\ 
     239 ,0  &   9,     38     & 0.113 &   447    &    4401 &      10 \\ 
     249 ,0  &   16,      29   & 0.107 &     600  &      6150 &      10 \\ 
     313 ,0  &   48,     -26   & 0.096 &     756  &      10754 &      7 \\ 
     315 ,0  &   49,     -28   & 0.081 &     498  &      9101 &      5 \\ 
     317 ,0  &   50,     -30   & 0.097 &     235  &      7614 &      3 \\ 
     319 ,0  &   51,     -31   & 0.081 &     527  &      6410 &      8 \\ 
     323 ,0  &   54,     -35   & 0.048 &     194  &      4732 &      4 \\ 
     333 ,0  &   62,     -43   & 0.069 &     264  &      2477 &      10 \\ 
     339 ,0  &   67,     -47   & 0.063 &     273  &       1680  &      16 \\ 
\hline 
\label{Table_fields} 
\end{tabular} 
\end{minipage} 

\end{table*}

Iron abundances were derived from medium resolution spectra (R=6000)
covering 3700\AA \, to 4700\AA~ using a methodology based originally
on that developed by \citet{b103} and extended by WGN. The technique,
which uses the CaK-line to determine iron abundance, is calibrated
using halo and thick disk field stars. We measured abundances only for
spectra having more than 100 net counts per 0.34\AA~ pixel at 4150\AA.  The
typical median counts for the sample is ~250 per 0.34\AA~ pixel at 4150\AA.
As noted below, this leads to an abundance uncertainty of
$\Delta$[Fe/H] $\sim$ 0.2 dex.  Radial velocities were determined by cross
correlation against sky and standard stars, providing velocities with
random plus systematic accuracy better than 10 km/s.

The selection function employed in obtaining the data was as follows:
\begin{center} 
$17.0 < g < 18.5$ \\ 
$-1.0 < (g - r)_0 < 0.5$\\ 
$-0.5 < (r - i)_0 < 0.6$\\ 
$(u - g)_0 \ge 0.6$\\ 
 \end{center} 
 This selection function preferentially selects thick disk and halo
stars, mainly F/G dwarfs, and minimises contamination from the thin
disk population. In addition, in several fields, fainter halo turnoff
candidate stars were targeted to exploit periods of very good seeing, together with Blue Horizontal Branch stars. This
extra selection was 
\begin{center} 
$18.5 < g < 19.5$ \\
$(g-r)_0 < 0.25$ \\
\end{center} 

although in practise few stars in this faint extension delivered spectra
of sufficient signal-to-noise ratio to deliver abundances, and essentially
none with $g>19.0$.

Figure \ref{SDSScmd}(a) shows the colour-magnitude diagram (CMD) for
all stars in the region $133^{\circ} < \mathrm{RA} < 339^{\circ}$ and
$-1^{\circ} < \mathrm{Dec} < 1^{\circ}$ from the
SDSS Data Release 7 \citep{b102}. The group of stars centred
around $(g-r)_0$ = 1.3 and faint apparent magnitudes are intrinsically
faint, nearby, low mass stars. These, from their kinematics and
metallicity estimates, are identified to be thin disk stars. The stars
centred around $(g-r)_0$ = 0.4 are older populations of stars with a
bluer main sequence turn-off. There is a shift at about $g_0$ = 18.5 in
the average colour of the stars, and this is caused due to the
different metallicities of the two groups of stars. The stars with
brighter apparent magnitudes have a relatively redder turn off, due to
relatively higher metallicity - this is the thick disk population. The
stars with bluer turn-off have comparably old ages, but lower
metallicity compared to the thick disk, and are halo stars.
 
Figure \ref{SDSScmd}(b) illustrates how the CMD changes when the $(r-i)_0$
and $(u-g)_0$ colour cuts are imposed on \ref{SDSScmd}(a). The number of
stars is reduced to 75\% of the original sample. The $(r-i)_0$ colour-cut
preferentially picks out the thick disk and halo stars, over the thin
disk ones, while the $(u-g)_0$ colour cut eliminates most quasars. This fact is
well illustrated by the colour-colour diagrams in Figure
\ref{SDSScmd}(c) and (d). The black squares in \ref{SDSScmd}(b) mark
out the $g$ and $(g-r)_0$ colour cuts imposed in this WGN AAOmega
Survey. This clearly shows that the selection function achieved the
goal of preferentially identifying thick disk and halo stars, while
minimising contamination from the thin disk and quasars. The F/G
dwarfs thus selected are at distances of a few  kpc.
 
Data Release 4 (DR4) of SDSS served as the
original input catalog to the AAOmega Survey. Thus the astrometric and
photometric information of the WGN selection function is from SDSS
DR4. For this work, the photometry was updated to the more recent Data
Release 7 (SDSS DR7).
 
The stars were cross-matched between DR4 and DR7 based on their unique
SDSS identifier. Astrometry and photometry compare well between SDSS
DR7 and DR4.  The positions of stars from the two datasets are
consistent, as expected. The extinction corrected apparent magnitudes
from DR4 are displaced from the corresponding values from SDSS DR7 by
0.002 magnitude with a $1\sigma$ spread of 0.01 magnitude, which is
negligible. Similarly so for the $g-r$ colour. This suggests that the
recalibration of photometry/extinction effects in DR7 as compared to
DR4 is minimal. 

Figure \ref{spectrophoto_compare} shows the comparison between the
spectroscopic results of WGN (obtained from measurements in the
AAOmega survey) for the 84 stars in common with the spectroscopic
observations released as part of SDSS DR7. Metallicity values between
the two surveys are in very good agreement: they are offset in the
mean by less than about 0.1 dex and show a scatter of 0.2 dex at the
$1\sigma$ level. Radial velocities from the two surveys are also in
good agreement, with a mean offset less than 2 km/s ($1\sigma = 3$
km/s). (The few large outliers are likely to be due to variable
stars). Both the metallicity and velocity offsets indicate
uncertainties we will adopt for the AAOmega survey.

We thus have available for analysis 32 fields spanning the equatorial
stripe, each some 2 degree squared area, centred on the listed
line of sight coordinates, with [Fe/H] and radial velocities from WGN
and $ugriz$ photometry and positions from SDSS DR7.
 
\begin{figure*}
\begin{minipage}{180mm} 
\begin{center} 
\includegraphics[width=0.7\linewidth]{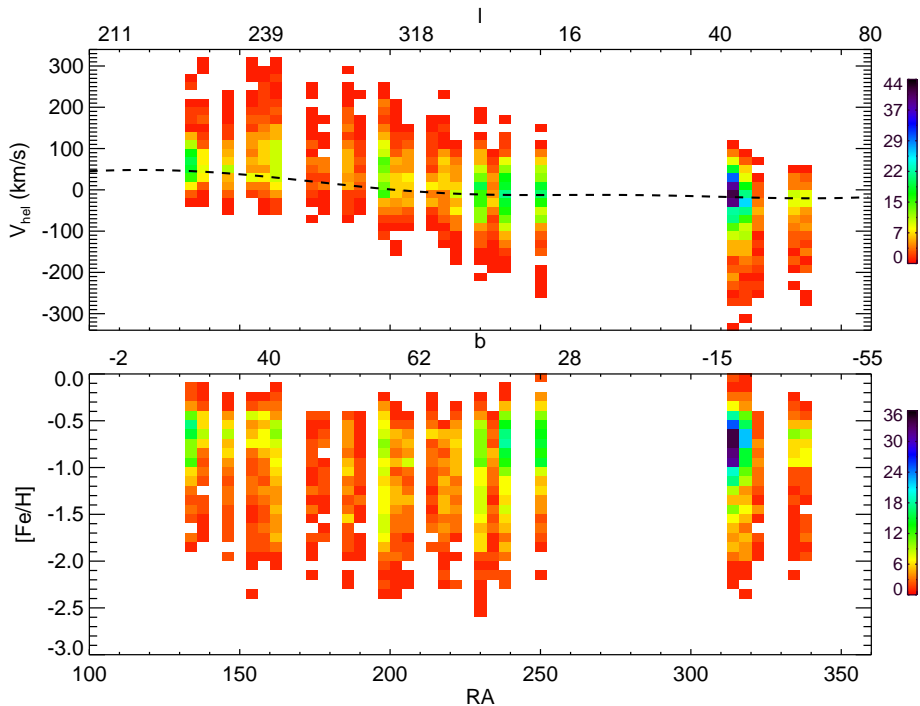} 
 \caption{Measured heliocentric velocity and metallicity as a function
   of RA from the WGN dataset analysed here. The corresponding
   Galactic longitude $l$ and latitude $b$ of the fields is shown in
   parallel axes. RA, $l$ and $b$ are quoted in degrees. The colours
   indicate the number of stars within each pixel (colour-coded in the
   colour bar) going from red for small to blue for large star
   count. The black dashed line shows an azimuthal velocity of 180
   km/s at a distance of 2  kpc projected onto the line of sight and is
   over-plotted to guide the eye along the thick disk population.}
\label{VRvsRA} 
\end{center} 
\end{minipage} 
 \end{figure*}

Figure \ref{VRvsRA} presents an overview of the spectroscopic
measurements from the WGN survey. It shows a plot of the measured
heliocentric velocity and metallicity vs RA. This broadly indicates
the presence of both the non-rotating, metal-poor halo component and
the rotating and relatively metal-rich thick disk component. An
azimuthal velocity of 180 km/s at a distance of 2  kpc projected onto
the line of sight is over-plotted for reference.

\section[Properties and Calibration of the WGN sample]{Properties and Calibration of the WGN sample} 
\subsection{Completeness Correction} 
 
Completeness is always a significant issue with any spectroscopic
dataset as there are many more stars which are present in a given
direction of the sky than the number for which we have spectra. With
the introduction of multi-fibre spectrographs in the last decade,
spectra of multiple objects are now obtained simultaneously but this is
still restricted to a few hundred objects per field of view.

There is a further bias towards brighter stars which arises from
discarding those spectra from the sample which have low
signal-to-noise ratio. The sample for which spectrometric parameters
are derivable has an additional bias against metal-poor, hot, blue
stars, which generates biased completeness in the purely photometric
input selection function, described in Section 2.1.

\subsubsection{Completeness of SDSS DR7}
We first check the completeness of SDSS by comparing to a deeper dataset, namely COMBO-17. COMBO-17 is a spectrophotometric survey which imaged 5 fields at high Galactic latitudes covering 1 square degree of the sky in 17 optical filters. It obtained deep and sharp r-band data and the selection of stars is complete down to $r \sim 23$ \citep{b23}

The SDSS DR7 photometry is sufficiently close to 100\% complete in the magnitude
range $16.0 < r < 21.0$ (which encompasses the range of the WGN
dataset $16.5<r<19.0$), in comparison to overlapping fields in the
COMBO-17 survey. \footnote{For the comparison of completeness between SDSS
  and COMBO-17 see {http://www.sdss3.org/dr8/imaging/other\_info.php}
}. 

It is thus reasonable to adopt SDSS DR7 as complete in the magnitude range of relevance to WGN. We therefore go on to use SDSS DR7 to evaluate the completeness of the WGN sample.

\subsubsection{Completeness of WGN}
We find that the WGN dataset has completeness levels that are as high
as 25\% of all potential candidates having spectra of sufficient
quality to allow derivation of both [Fe/H] and radial velocity. The
last 2 columns of Table \ref{Table_fields} show the number of stars in
the given field in the SDSS master photometric catalog (employing the
same selection functions as in the data) and the relative completeness
of the WGN data field-by-field.

\subsubsection{Weighting Method} 
 
We describe here a weighting method in colour-magnitude space to
correct the WGN spectroscopic sample for completeness inside the WGN
selection function. This method is widely applied in sampling, in
situations as diverse as opinion polling and ornithology. We note its
applications in studies of the thick disk in astronomy below.

Stars in a colour-magnitude bin of the spectroscopic data are weighted
up to the actual number of stars available for observation in that bin,
by a weighting factor $W$ which is obtained by taking the ratio of the
density function of stars in colour magnitude space $\mathrm{D[g_0,(g-r)_0]}$ of the
SDSS DR7 catalogue to the data themselves:
 
\begin{equation} 
W= \frac{\mathrm{D_{SDSS DR7}[g_0,(g-r)_0]}}{\mathrm{D_{WGN}[g_0,(g-r)_0]}} 
\end{equation}

Three different methods of computing the density function for both the
Sloan photometric superset and the WGN dataset were tried : (a)
probability density functions were calculated by convolving each data
point with a Gaussian kernel function; (b) one dimensional number
density distributions (as a function of apparent magnitude) for
individual colour bins were determined by simple polynomial fits to 1D
apparent magnitude histograms; (c) a two-dimensional histogram
(i.e. number of stars as a function of colour-magnitude) was used to
represent the distribution of stars in the colour-magnitude plane.
 
The three methods provide progressively decreasing degrees of
smoothing ((a)-most smoothing, (c)-least smoothing) across the
colour-magnitude plane.

The pixel or bin choices in each of these methods was made so that we
sufficiently resolve the main features of interest on the CMD (i.e. the
different Galactic stellar populations), while ensuring that the
pixels contain sufficient number of stars relative to the noise
level. This choice may also be dependent on line-of-sight. 

We found that the distribution functions produced by these three
methods did not differ significantly, indicating that the particular
method chosen to compute the density distributions in colour magnitude
space is not crucial. We chose method (c) as the most robust of the
three methods for the rest of the analysis. Along with best
performance with low number statistics and ease of computation, this
method has particular benefits over the other two.  Method (a)
involves the risk of the smearing of tiny features that may be of
importance through over-smoothing, while edge effects introduce
problems in the polynomial fitting technique of method (b).

\begin{figure*} 
\begin{minipage}{180mm} 
\begin{center} 
\includegraphics[width=0.7\linewidth]{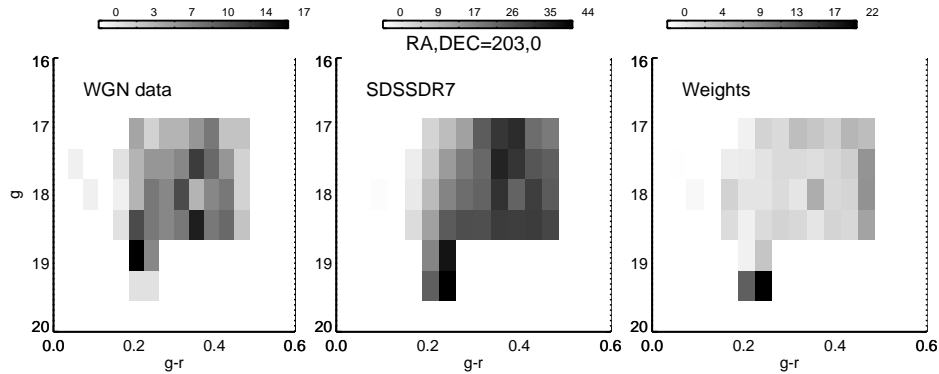} 
 \caption{Completeness correction is implemented for the WGN data
   through a weighting method. This method is illustrated here for the
   (not untypical) field with coordinates RA,Dec = $203^{\circ},
   0^{\circ}$. Left: the distribution in colour magnitude space of the
   stars observed by WGN in the 2x2 degrees field. Middle: SDSS DR7
   data for a 6x6 degrees region centred around the same coordinates,
   for the same colour-magnitude cuts. Right: the corresponding
   weights, which are calculated as the ratio of the middle to the
   left-most panel, multiplied by the area factor, which is given by
   the area covered on the sky by the left-most panel divided by the
   area covered by the middle panel}
\label{weighting_example} 
\end{center} 
\end{minipage} 
 \end{figure*} 

This weighting function was computed field by field, taking into
account that extinction varies across the sky and that the form of the
CMD also varies as an effect of the different density profiles of the
halo and thick disk. The complete photometric sample from SDSS DR7 was
selected for the same colour magnitude cuts, but for a larger spatial
region (6x6 degrees) centred on each field, to minimise
noise. Colour-magnitude bins of size $\Delta g,\Delta (g-r) =
0.7,0.13$ mag were found to be an appropriate choice for all
fields. The weighting method is illustrated in Figure
\ref{weighting_example} for a single field, centred at RA,Dec=
203$^{\circ}$,0$^{\circ}$. The first panel shows the WGN data for a
single 2dF field (i.e. a circular field of radius 1 degree centred on
the mentioned coordinates).  The middle panel shows the SDSS DR7 data
for a larger region centred around the same coordinates, for the same
colour-magnitude cuts. The last panel shows the weights, which are
calculated as the ratio of the counts in the middle to those in the
left-most panel, multiplied by an area factor (given by the area
covered on the sky by the left-most panel divided by the area covered
by the middle panel).  In order to avoid large errors from low
signal-to-noise pixels, the distribution of weights in every field is
cut off at the 90th percentile. Objects having weights greater than
the 90th percentile value are reassigned to this new maximum value.
 
\begin{figure} 
\includegraphics[width=1.0\linewidth]{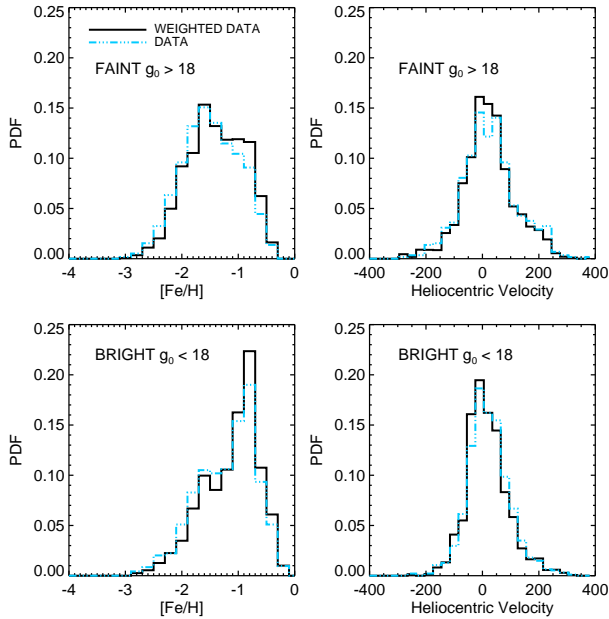} 
 \caption{Observed (dashed blue) and Completeness Corrected (solid black)
   Distribution Functions of [Fe/H] and radial velocity for the WGN
   data fields within $190^{\circ}<$RA$<220^{\circ}$. This covers the
   high latitude region $54^{\circ}<b<63^{\circ}$, in the inner Galaxy
   $312^{\circ}<l<347^{\circ}$. The data has been split into fainter
   stars $g_0 > 18$ in the top panels and brighter stars $g_0 < 18$ in
   the bottom panels. The unweighted metallicity distributions show a
   slight bias to a higher number of metal-poor stars,particularly in
   the faint bins, which are likely to be predominantly halo stars
   given these are high latitude fields. Since these fields are
   towards the inner Galaxy, metal-rich thick disk stars must also be
   numerous, and the weighting method corrects for this, seen more
   evidently in the faint bin.  The velocity distributions are in
   general smoothed by the completeness correction.}
\label{HISTOS} 
\end{figure}

In Figure \ref{HISTOS} we illustrate how the metallicity and velocity
probability distribution functions (PDF) change when the completeness
correction is applied. The histograms shown are for the high-latitude
fields, with RA from 190$^{\circ}$ to 220$^{\circ}$. This covers 6
fields at high latitudes ($54^{\circ} < b < 63^{\circ}$) in the inner
galaxy ($312^{\circ} < l < 347^{\circ}$). The data have been split
into faint ($g_0 > 18$, top panels) and bright ($g_0 < 18$, bottom
panels) stars. In both cases, the unweighted metallicity distributions
show a high number of metal-poor stars, particularly in the faint
bins, which are likely to be predominantly halo stars given these are
high latitude fields.  Since we are looking towards the inner galaxy,
metal-rich thick disk stars must also be numerous. Applying the
weights corrects the distribution for the fact that the data do not
show high enough normalisation at the metal-rich end, the effect being
more evident in the faint bin. We note that the heliocentric velocity
distribution is centred on zero as one would expect for the chosen
line-of-sight towards the inner Galaxy. We can see that the velocity
distributions are in general smoothed by the completeness correction.

\subsubsection{Errors on the Weights}  

The error on the weight $W$ was calculated as
\begin{equation} 
\sigma_w[\mathrm{g_0,(g-r)_0}] = \frac{W}{\sqrt{\mathrm{D_{WGN}[g_0,(g-r)_0]}}} 
\end{equation} 
  
\begin{figure} 
\begin{center} 
\includegraphics[width=0.6\linewidth]{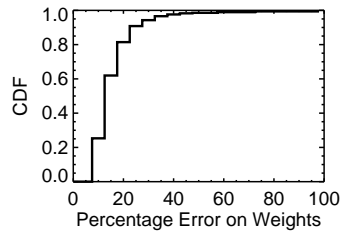} 
 \caption{The cumulative distribution function of the percentage
   errors on the weights assigned to the stars by our weighting method
   for completeness correction. More than 80\% of the stars have less
   than 20\% error, confirming that the completeness corrections
   derived here are accurate. }
\label{Weight_Err} 
\end{center} 
\end{figure}

Figure \ref{Weight_Err} shows the cumulative distribution function of
percentage errors on the weights. This shows that about 80\% of the
stars have less than 20\% error on their weights. In all further
analysis, objects with weight error above 50\% were
discarded. This eliminated 164 objects from the
    sample, only 90 of which have $(g-r)_0 > 0.2$ and thus belong to the region of CMD being analysed. Thus by cleaning just 1\%
    of the sample, we remove errant weights without introducing any
    additional bias to our sample .

\subsubsection{Advantages of the adopted weighting method} 

Weighting the observed data to allow for parameter-dependent
incompleteness is a standard requirement in statistical sampling, with
many types of approaches. The method we apply here is both very
obvious and very simple, and so widely used. To our knowledge it was
first applied in determination of the thick disk abundance
distribution in section 3.3 of \citet{b7}, but there may be earlier
examples.

In some other investigations (for example, \citet{b16}), completeness
correction is done by comparing the dataset in question with a
classical power law, which represents a smooth fit to the number
distribution of stars as a function of apparent magnitude, or by
comparison to the stellar distribution (in colour-magnitude space)
predicted by a Galaxy model, which might also appropriately take into
account a model of extinction. The weighting method used here
also uses the additional parameter of colour. There is a definite
variation of the value of the weights along the colour axis for a
given magnitude which makes it important to take into account this
additional parameter.

The advantage of this adopted method of accounting for completeness is 
that it is expected to preserve information from any interesting
groups that may exist in velocity-metallicity space, without being
limited by potential mismatches between a model and the actual star
counts.

 Similar methods of completeness correction
    have been used by some recent papers analysing stars from the Sloan Extension for Galactic Understanding and Exploration survey (SEGUE). \citet{Cheng2} employ a similar method of weighting in
    $g-r$ vs $g$ space to account for the incompleteness in old MSTO
    stars with spectroscopy in SEGUE's  low galactic latitudes
    fields. They multiply this weight with additional weights - one
    based on a luminosity function, to reduce bias against redder
    metal-rich stars in their sample selection; this was found to be
    mostly constant, and another based on area covered on
    the sky to account for regions with high extinction that were not
    considered for spectroscopy; this does not apply to the WGN
    data. \citet{LV12} analyse G type SEGUE stars and employ weights
    which are a function of $u-g$,$g-r$ colours and distance. Their
    distances are determined using relations from \citet{b13} and though they state
    that uncertainties in distance determination will not affect the
    weights, using apparent magnitude instead avoids this
    potential problem.

\subsubsection{Comparison to Photometric Metallicities} 
\citet{b13} describe a method of estimating metallicities for those
F/G main-sequence stars for which spectra have not been obtained in
SDSS, from their $u-g$ and $g-r$ colours, to achieve the same goal as
above. Building on the classical photometric metallicity calibrations
applied to Galactic structure by \citet{PMref1}, \citet{PMref2},
\citet{PMref3} and \citet{PMref4}, they first use a sample of SDSS
stars with spectra to calibrate the metallicities and thereafter use
this relation to estimate the metallicities of the complete sample of
stars within the same region. The photometric metallicity was
calibrated using SDSS DR6 data in the region $0.2 < g - r < 0.6$, $14
< g < 19.5$. The prescription was improved in a follow-up paper by
\citet{b3} (hereafter MWIII) replacing the DR6 data by DR7. Their
calibration is expected to produce good results in the metallicity
range $-2.2<\mathrm{[Fe/H]}<-0.2$; at the lower limit, [Fe/H]
asymptotes due to insensitivity of u-g colour to decrease in
metallicity, while at the upper end there is a systematic
underestimation of the [Fe/H] values, possibly due to a lack of metal
rich stars in the calibration data.
 
\begin{figure} 
    \includegraphics[width=1.0\linewidth]{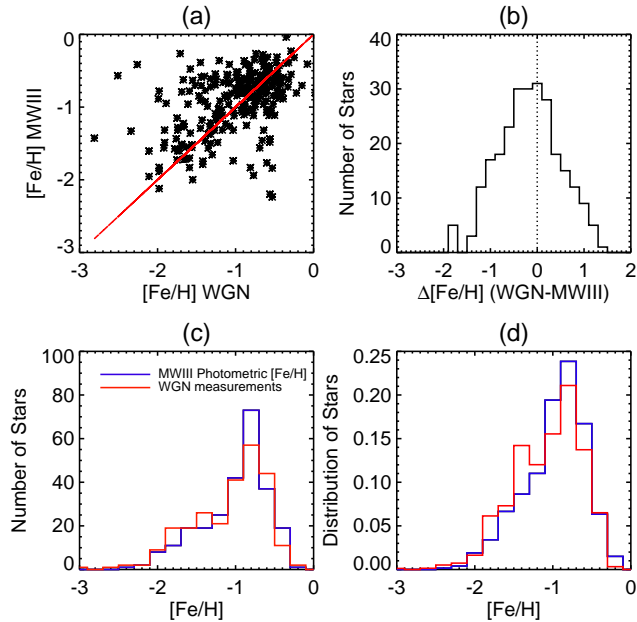} 
 \caption{Comparison of WGN metallicities to photometric metallicities
   from \citet{b3} (referred to as MWIII in the plots) for the sample
   field RA,Dec=145$^{\circ}$,0$^{\circ}$. [a] Comparing the measured
   WGN [Fe/H] to the [Fe/H] calculated for these stars by the
   photometric metallicity method. The red line shows a one-to-one
   correlation. [b] Difference between the WGN [Fe/H] and the
   photometric [Fe/H]. The photometric metallicities have a mean
   offset of $\sim0.16$ dex towards higher metallicities with respect
   to WGN measurements. The $1\sigma$ scatter is 0.68 dex.  [c] The
   observed metallicity distribution function (MDF) from the WGN data
   compared to the MDF obtained by calculating photometric
   metallicities for the same stars. [d] Completeness corrected WGN
   MDF compared to the MDF resulting from calculating the photometric
   metallicities for the complete SDSS photometric sample in that
   field.}
\label{MWIII} 
\end{figure}

Figure \ref{MWIII} shows a comparison between the metallicity
distribution functions resulting from WGN metallicities and the
photometric metallicities of MWIII for the sample field
RA,Dec=145$^{\circ}$,0$^{\circ}$. Figure \ref{MWIII}(a) shows MWIII
estimates for those stars which have metallicities directly measured
in the WGN survey. We can see that at both high and low metallicities
the MWIII method shows large deviations from the measured
values. Figure \ref{MWIII}(b) more clearly illustrates that MWIII
shows systematics of 0.16 dex towards higher metallicities, and very
large random errors.  The bottom panels show how the metallicity
distribution functions (MDF) get affected as a result. While Figure
\ref{MWIII}(c) shows a comparison between measured WGN metallicities
and corresponding photometric metallicities for the same objects,
Figure \ref{MWIII}(d) compares the weighted distribution function in
the same field with the photometric metallicities for the complete
SDSS photometric sample in that field. Even though the distributions
produced by the two methods do not differ very significantly, the
star-by-star comparison shows us that the wings of the MWIII
distributions are dominated by large calibration errors. Photometric
metallicities, unsurprisingly, blur out the metallicity distribution
functions. The photometric MDF in Figure \ref{MWIII}(d) looks smooth,
while the underlying spectroscopic MDF in the same line of sight
suggests a much more discrete halo plus thick disk two-component MDF.
 
The weighted dataset that we obtain as a result is thus a
completeness corrected representation of the WGN selection function.
 
\subsection{Distance Estimates} 

We now have a completeness corrected dataset of thick disk-halo stars
containing information about each star's position on the sky and
radial velocity apart from $ugriz$ photometry and metallicity. In
order to use this information to analyse the full phase-space
structure of the Galaxy we need an estimate of the third position
coordinate - distance. Good stellar distance estimates are key to
making correct inferences about global Galactic structure.

For Galactic distance scales larger than a few hundred parsec, while
we await Gaia, methods of spectroscopic parallax are commonly used to
estimate distances. Information on the temperature and metallicity of
the stars is used along with stellar evolution models to make an
inference about the star's absolute magnitude and hence deduce its
distance given the measured brightness.

In this work we use a $\chi^2$ technique to fit the data to the
main sequence of theoretical stellar isochrones in order to estimate
distances to the stars. In the following sections we first outline
this method, thereafter describing the use of star cluster data to
choose the best stellar evolution models, and to test the performance of
the method. The errors on the estimated distances are analysed and
calibrated. We finally present the distance estimation for the WGN
data in Section 3.2.3. and verify the robustness of these estimates by
including the additional constraint of infrared data and comparing to
estimates from photometric parallax.

\subsubsection{Method} 
Distances are estimated by a $\chi^2$ fitting of the data to the main
sequence of theoretical stellar isochrones. We start with a grid of
isochrones with metallicities in the range -2.5 to 0.0 dex such that
the choice of the metallicity grid evenly samples colour-magnitude
space. The ages of the isochrones are restricted to greater than 10
GYr, since our colour-cut selection ensures that only stars from the
old Galactic stellar populations are preferentially selected (the thick
disk is estimated to have an age $ \sim 12$ GYrs \citep{b7}, while the
halo is similarly old with a most recent estimate of inner halo field
stars to be $ \sim 11.4 \pm 0.7$ GYrs \citep{b11}).

For each set of photometric and metallicity data describing a star,
the theoretical isochrones of closest metallicity are first
chosen. We then consider those model stars on the chosen metallicity
isochrone which are within an n$\sigma$ colour ellipsoid of the data
point. The n$\sigma$ bracket was set to optimise the number of stars
for which distances are determined.

Let data point $j$ have $k$ model matches within an n$\sigma$ colour
ellipsoid. These $k$ matches have the same metallicity (by choice of
isochrone) and have varying ages greater than 10 GYrs. The $\chi^2$
value for a data point $j$ with respect to any such model point $k$
is then calculated as follows :

\begin{equation} 
(\chi^2)_{model_k}^{j}= \sum_{x} ({\frac{x_{data_j} - x_{model_k}}{\sigma_{x_{data_j}}}})^2 
\end{equation}  
where the index $x=$ $\lbrace u-g,g-r$,$r-i,i-z\rbrace$,
  each model match k having the associated probability given by,

\begin{eqnarray} 
\nonumber P((\chi^2)_{model_k}^j)= \frac{ (\chi^2)^{\frac{m}{2}-1} exp(-\frac{\chi^2}{2})}{2^{\frac{m}{2}} \Gamma(\frac{m}{2})}
\end{eqnarray} 
where $m$ is the number of degrees of freedom. The luminosity
functions appropriate to the isochrones are then used to evaluate the
probability $P(\mathrm{M}_k|\mathrm{age}_k)$ of a star having an
absolute magnitude $\mathrm{M}_k$, as suggested by each model match
$k$, given its age.

The probability assigned to each model star $k$ of being the best
match to the data point $j$ is then given by
\begin{equation} 
P_{model_k}^j=P((\chi^2)_{model_k}^j)P^j(\mathrm{M}_k|\mathrm{age}_k)
\end{equation} 

The closest model match to the star $j$ is then found by maximising
$P_{model_k}^j$, and from this absolute magnitude the distance to the
star is inferred.

\subsubsection{Testing the distance determination method} 

We use star cluster data with well-known distance estimates to select
the set of stellar evolution models which are most consistent with the
SDSS photometric data on those clusters, and thereafter to analyse the
performance of the method.

\paragraph{Star Cluster Data} 
\citet[][hereafter An08]{b1} present photometry for stars in 17
globular clusters and 3 open clusters in a value-added catalog of
SDSS, derived by reducing SDSS imaging data in crowded cluster fields
using the DAOPHOT/ALLFRAME suite of programs. Distances were derived
to 6 of these globular clusters and 2 of the open clusters in a
subsequent paper by \citet[][hereafter An09]{b2} using an isochrone
fitting technique in the $ugriz$ filter. This reproduced previously
available estimates calibrated from Hipparcos parallaxes in the
Johnson-Cousins photometric system.
 
We applied our method to individual stars in five globular clusters
looked at by An09 - M5, M3, M13, M92 and M15 ranging in metallicity
from -1.26 to -2.42 dex and all older than 12 GYrs - and compared our
resulting distance estimates. All these globular clusters have
photometry with dependable zero point error from An08. They have
[Fe/H] estimates from \citet{b10} based on high resolution spectra. We
also chose two open clusters from An08 to cover the metal-rich end of
the WGN [Fe/H] range - M67 and NGC2420 with [Fe/H] = 0.00 and -0.37
dex respectively. The data for NGC2420 was presented in An08 but not
analysed by An09. Its distance, metallicity and reddening values were
adopted by An08 from \citet{AT06}.

 Cluster members were chosen based on a good image quality index, a
 measured characteristic radius close to that of the chosen PSF in
 pixels, and by removing pixels with either no detection or saturated
 detection.  The results are discussed below, after we consider the
 optimal choice of isochrones.

\paragraph{Selection of isochrones}

\begin{figure*} 
\begin{center} 
\includegraphics[width=0.8\linewidth]{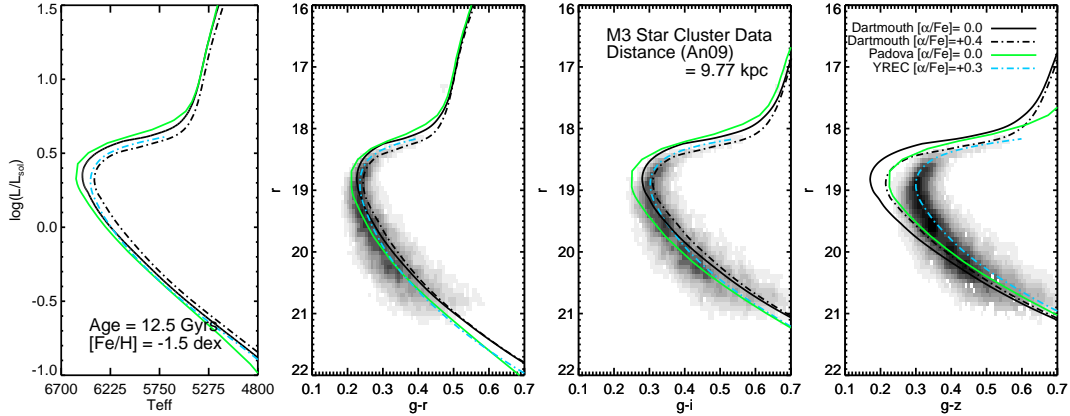} 
\caption{We compare stellar evolution models to star cluster
  data. Three sets of isochrones - Dartmouth, YREC and Padova - are
  compared with the SDSS photometric data for the globular cluster
  M3. The adopted properties of M3 are: [Fe/H] = -1.50, age = $13.3
  \pm 1.4$ GYrs, distance \citep{b2} $9.44 \pm 1.03$ kpc. All
  isochrones shown have [Fe/H] = -1.5 and age =12.5 GYrs (the closest
  age to that of M3 which has isochrones in all three models). The
  YREC isochrone has $[\alpha/Fe]=+0.3$; two Dartmouth isochrones with
  $[\alpha/Fe]=0$ and $+0.4$ are shown; the Padova isochrone has
  $[\alpha/Fe]=0$. The leftmost panel plots LogL vs Teff, and this
  shows that all three models are in good agreement with each other in
  the luminosity region of relevance here.  The different approaches
  to convert luminosities and effective temperatures to magnitudes and
  SDSS colours are reflected in the next three panels, where the
  isochrones have been over-plotted with the SDSS photometric data for
  M3.  From this comparison we adopt the Dartmouth isochrones
  restricting their application to $g-r$ and $g-i$ colours.  }
\label{GC} 
\end{center} 
\end{figure*}

Several sets of isochrones are available matched to old stars with
SDSS photometry. An09, whose distance scale we compare to our own,
apply the Yale Rotating Evolutionary Code (YREC; \citet{Sills2000})
isochrones to estimate the star cluster distances. They discount the
$u$ filter as the models overestimate the flux in the $u$ bandpass
causing a 5\% difference between data and model for colours involving
this bandpass. More encouragingly, they find that the YREC model
colours in $g-r$,$g-i$ and $g-z$ agree well with the main sequence of
their globular cluster data. They also adopt an alpha-enhancement
scheme motivated by the observed amounts of alpha elements in field
and cluster stars.

In Figure \ref{GC} we compare three sets of isochrones - Dartmouth 
\citep{b5}, YREC and Padova \citep{b6} - with data for the Globular
Cluster M3. M3 has [Fe/H] = -1.50, estimated age = $13.3 \pm 1.4$ GYrs
and estimated distance $9.44 \pm 1.03$ kpc (An09). All isochrones
shown have [Fe/H] = -1.5 and age =12.5 GYrs (the closest age to that
estimated for M3 for which all three models have calculated
isochrones). An09 use the YREC isochrone with
$\mathrm{[\alpha/Fe]}=+0.3$. The Dartmouth database provides
isochrones with a coarse grid in alpha-enhancement, from which we pick
two isochrones of $\mathrm{[\alpha/Fe]}=0$ and $+0.4$.  The Padova
database does not provide isochrones with alpha-enhancement, so we
pick $\mathrm{[\alpha/Fe]}=0$.

The leftmost panel of Figure \ref{GC} plots log luminosity vs
effective temperature, and this shows that all three models are in
good agreement with each other, with the exception of the systematic
deviation of the Padova isochrones on the lower main-sequence, which
is not relevant here. The three isochrone groups employ different
methods to convert luminosities and effective temperatures to
magnitudes and colours, which is reflected in the next three panels of
Figure \ref{GC}. Here the isochrones have been over-plotted with the
data for M3, assuming the distance to M3 is as determined in An09
(i.e. $\sim 9.77$ kpc).

In the Dartmouth Stellar Evolution Model, the model colours suffer
from inaccuracy in synthetic fluxes for central wavelengths bluer than
$\sim5000$ \AA~ thus making the $u$ filter very uncertain.
Additionally, problems exist in the underlying SDSS calibration of the
$u$ and $z$ filters, which have larger statistical errors in their
zero points and poorer photometric uniformity (both by approximately
50\%) as compared to the $gri$
filters.\footnote{http://www.sdss.org/dr7/algorithms/fluxcal.html}
From Figure \ref{GC} we see that while the Dartmouth isochrones seem
to agree with data in the $g-r$ and $g-i$ colours, there is a large
deviation (almost 0.5 magnitude near the turn-off of the cluster) in
$g-z$. This suggests that there are larger errors associated with the
model $z$ filter, as compared to $gri$. The Padova isochrones seem to
have poorer agreement with data both in $g-i$ and $g-z$ colours.

In addition to good agreement to data in the $gri$ filters, the
Dartmouth database provide a finer grid of both metallicities (user
determined) and ages (0.5 GYrs). Both the Padova and YREC databases
have more sparse sampling at higher ages relevant to the WGN
data. Thus we adopt the Dartmouth isochrones, excluding the $u$ and
$z$ bandpasses from the isochrone fitting i.e. in Equation 3 we use
only colours $x= \lbrace g-r,r-i\rbrace$.

\paragraph{Performance}
The distance determination method was tested on the chosen star
clusters by picking the isochrone of the closest [Fe/H] to the
cluster. For all the globular clusters, the age of the isochrones was
restricted to above 12 GYrs. Matches within a 2$\sigma$ ellipse of two
colour filters $g-r$ and $r-i$ was used in the $\chi^2$ fitting. A
2$\sigma$ ellipsoid was sufficient to obtain distances for at least
80\% of the cluster data.

The performance of the distance determination method can be analysed
by means of an example case of M13 shown in Figure
\ref{GCs_performance_M13}. The leftmost panel shows a CMD of the
cluster containing all cluster members for which distances were
determined. The second panel from the left shows a plot of $g-r$ vs
the absolute $g$ magnitude assigned to the star by the distance
determination method. The third panel from left shows the deviation of
this distance estimate from the ``true'' (An09) distance of the cluster
as a function of apparent magnitude $g$. This panel has been colour
coded to highlight groups of stars which show extremely deviant
distances. These deviant groups are as follows:

\begin{figure*} 
\begin{center} 
\includegraphics[width=1.0\linewidth]{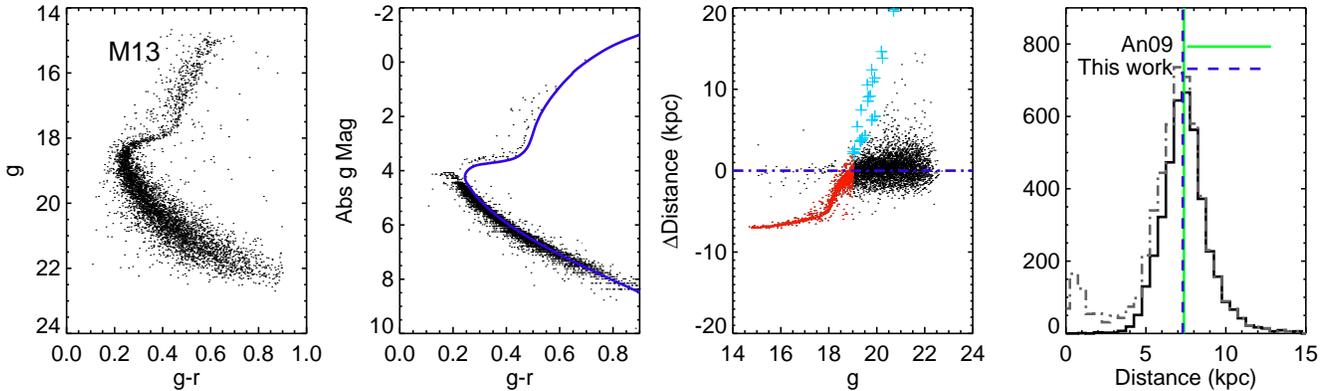} 
\caption{Results of our distance determination method applied to the globular
  cluster M13. The left most panel shows the CMD of the cluster, the
  second panel from left shows the absolute g magnitude recovered for
  each star from the isochrone fitting, the third panel from left
  shows the deviation of estimated distance from the An09 value as a
  function of apparent magnitude $g$ (red points mark giants
  mis-classified as main sequence stars; blue plus signs mark the main
  sequence stars which are mis-classified as giants) and the rightmost
  panel shows the distance distribution recovered by the method for
  all stars (grey dashed line) and for correctly classified main
  sequence stars (solid black line)}
\label{GCs_performance_M13} 
\end{center} 
\end{figure*}

\begin{figure*}
\begin{center}
 \subfigure[a]{%
            \label{}
		\includegraphics[width=0.8\linewidth]{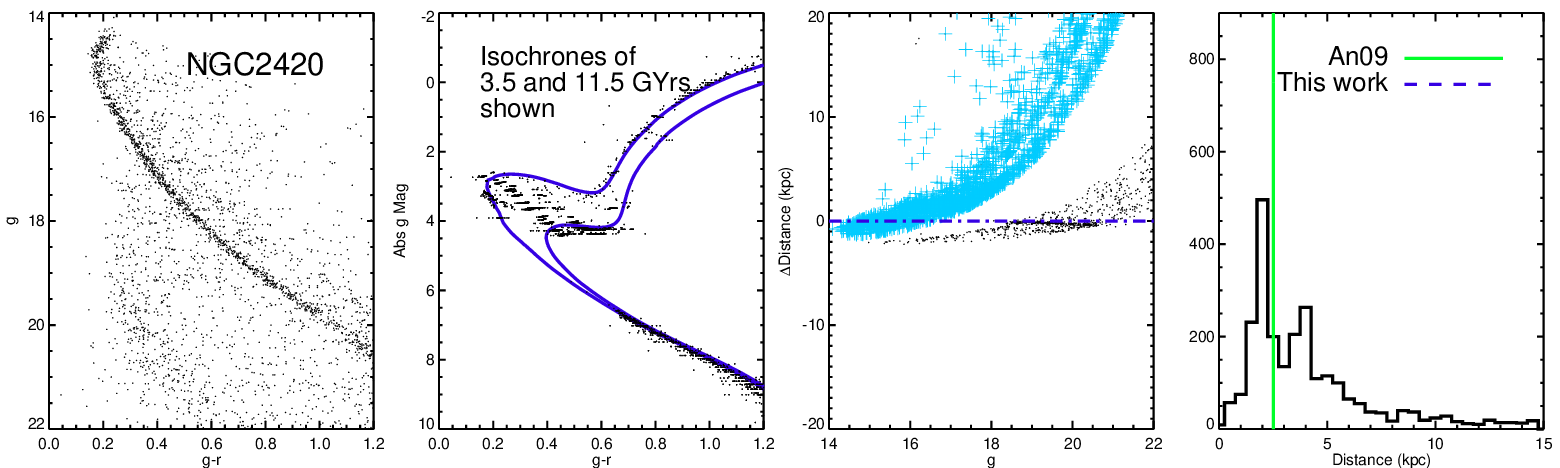}
        }\\
 \subfigure[b]{%
            \label{}
		\includegraphics[width=0.8\linewidth]{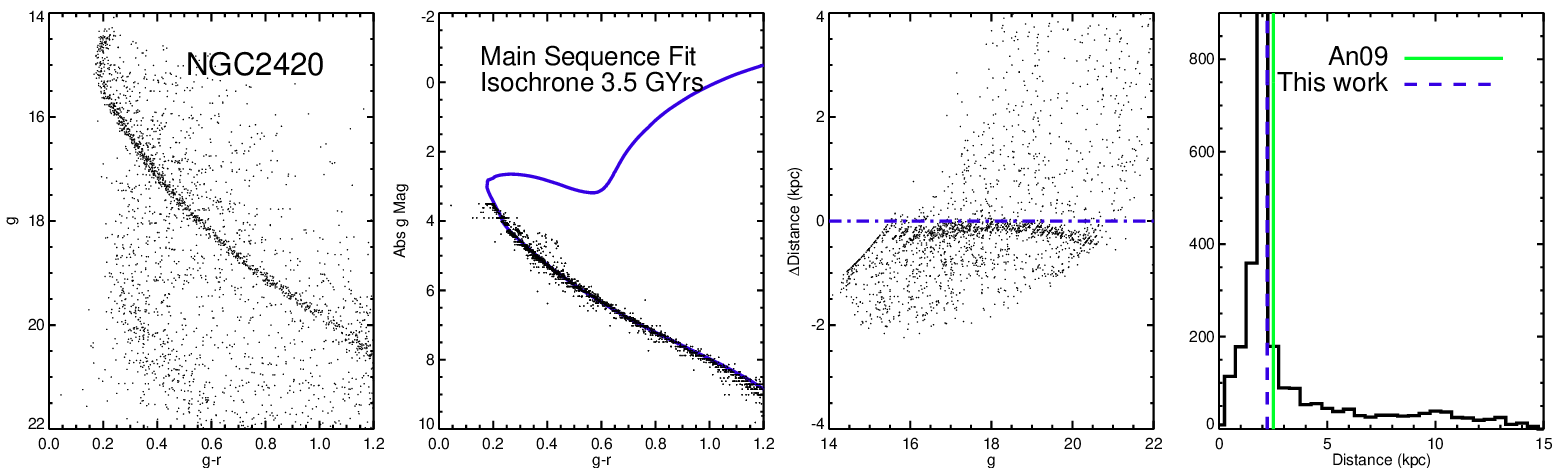}
        }%
\end{center}
\caption{Results of our distance determination method applied to the open
  cluster NGC 2420 ([Fe/H] = -0.37dex, Age $\sim$ 3.5 GYrs). (a) Top
  panels : Results on applying isochrone fit, restricting ages to
  above 3.5 GYrs. In this case almost all stars below $g-r = 0.6$ are
  incorrectly assigned to the sub-giant branch. (b) Bottom Panels :
  Results on applying isochrone fit, restricting ages to above 3.5
  GYrs and additionally restricting the fit to the main sequence
  region only. Here the problem of mis-classification of main sequence
  stars is resolved. In both (a) and (b), the left most panel shows
  the CMD of the cluster, the second panel from left shows the
  absolute g magnitude recovered for each star from the isochrone
  fitting, the third panel from left shows the deviation of estimated
  distance from the An09 value as a function of apparent magnitude $g$
  (blue plus signs mark the main sequence stars which are
  mis-classified as giants) and the rightmost panel shows the distance
  distribution recovered by the method for correctly classified main
  sequence stars.}
\label{OCs_performance_NGC2420}
\end{figure*}

\begin{itemize}

\item
The red points mark the giant stars (i.e stars having apparent
magnitude $g$ brighter than turn-off) but are incorrectly assigned to
be main sequence stars by the routine. As a result, their distances
are underestimated. These are isolated using a rough cut of $g < 19$
and absolute $g$ magnitude $>\sim 4$.

\item
The light-blue plus signs on the contrary are main sequence stars (i.e
stars having apparent magnitude g fainter than turn-off) that have
been incorrectly assigned to be giants, thus overestimating their
distances. These are isolated using a rough cut of absolute $g$
magnitude up to the value of 4.

\end{itemize}

Apart from these anomalies, good estimates of distances are assigned
to the bulk of the main sequence stars shown in black. This can be
seen more clearly from the rightmost panel in Figure
\ref{GCs_performance_M13} (solid black line) which shows the distance
distribution recovered for the cluster after eliminating all the
anomalous star groups (blue and red points) as described above. The
blue dashed vertical line which indicates the mean distance estimated
for the cluster by our method matches the green line which indicates
the value estimated by An09 within 5\%. The 1$\sigma$ spread of the
distribution is 1.1 kpc (placing $\sim$70\% of the stars within 15\%
of the mean distance).

In Figure \ref{OCs_performance_NGC2420}(a) we show the results of
testing the distance determination method at the metal rich end. All
panels are the same as in Figure \ref{GCs_performance_M13}. There are
no giants in the cluster, and hence no red points. The blue plus signs
mark out the main sequence stars mistakenly assigned to be giants, as
before. But the noticeable difference now is that \textit{all} main
sequence stars up to a $g-r$ colour of 0.6 (which is the relevant
$g-r$ colour range for the WGN data) seem to be mis-assigned as
giants. The reason for this can be understood from the form of the
Luminosity Function for high metallicities - for the higher
metallicities, the probability of finding a star near turn-off or on
the sub-giant branch is equal to (or more than) that of finding a star
on the Main Sequence, as against the lower metallicities where the
probability steadily decreases from lower main-sequence going towards
turn-off. Since we use the luminosity function as an additional
constraint to determine the absolute magnitude assigned to a star, all
main sequence stars are identified as sub-giants at high
metallicities.

Given the WGN data is comprised mainly of main sequence (mostly
turn-off) stars, we can remedy this problem by forcing the program to
choose model stars that only lie on the main sequence (placing a
cut-off at turn-off). The result of this is illustrated in Figure
\ref{OCs_performance_NGC2420}(b). We immediately see that all the
main-sequence stars are now assigned good distances and the recovered
mean distance is within 10\% of the An09 value.

While Figures \ref{GCs_performance_M13} and
\ref{OCs_performance_NGC2420} illustrate convincingly that the method
always recovers the correct distance for the cluster as a whole,
Figure \ref{OCs_performance_NGC2420}(b) also reveals that the mean
recovered distance could be a weak function of the photometry
itself. This trend is further discussed and calibrated in Section
3.2.2.4.

Our estimated cluster distances, together with those adopted for
comparison, are summarised in Table \ref{Table_distance}, along with
the various cluster parameters.  We see that our method reproduces
star cluster distances within 10\% of the distance scale adopted by
An09.

\paragraph{Result and Calibration of Errors}

\begin{table*}
 \begin{minipage}{140mm}
\begin{center}
  \caption{Star Clusters used to test the distance determination method}
  \begin{tabular}{@{}lccccccc@{}}
 \hline
Cluster & [Fe/H] & Distance        & E(B-V) & Age    & Dartmouth Isochrone & Result & \% Error\\
        &        & ( kpc)           &        & (GYrs) & [Fe/H] & Mean Distance &\\
 \hline
\hline
Globular Clusters \\
\hline
M5      & -1.26 & $7.11 \pm 0.20$  & 0.03 &  $12.2 \pm 1.3$ & -1.20 & 7.36 & 4\\
M3      & -1.50 & $9.77 \pm 0.32$  & 0.01 &  $13.3 \pm 1.4$ & -1.50 & 9.58 & 2\\
M13     & -1.60 & $7.37 \pm 0.18$  & 0.02 &  $14.3 \pm 1.1$ &  -1.60 & 7.30 & 1 \\
M92     & -2.38 & $8.55 \pm 0.15$  & 0.02 &  $14.4 \pm 0.9$ & -2.40 & 8.64 & 1\\
M15     & -2.42 & $10.47 \pm 0.55$ & 0.10 &  $13.5 \pm 2.5$ & -2.40 & 9.87 & 5\\
\hline
\hline
Open Clusters\\
\hline
M67     &  0.00 & 0.82             & 0.03 & 3.5             & 0.00 & 0.84 & 2\\
NGC2420 & -0.37 & 2.51             & 0.04 & 3.5             &  -0.40 & 2.23 & 11\\
\hline
\label{Table_distance}
\end{tabular}
\end{center}
\end{minipage}
\end{table*}

As noted in section 3.2.2.3, while the mean distance recovered for the
cluster by the distance determination method matches well with the
true distance, there is a weak trend of the recovered distance with
photometry. There is also an error which is introduced due to the finite [Fe/H] grid that
is chosen. In this section we analyse and calibrate these different
contributions to uncertainty in our distance estimates to individual
stars.

\begin{figure} 
\begin{center} 
\includegraphics[width=1.0\linewidth]{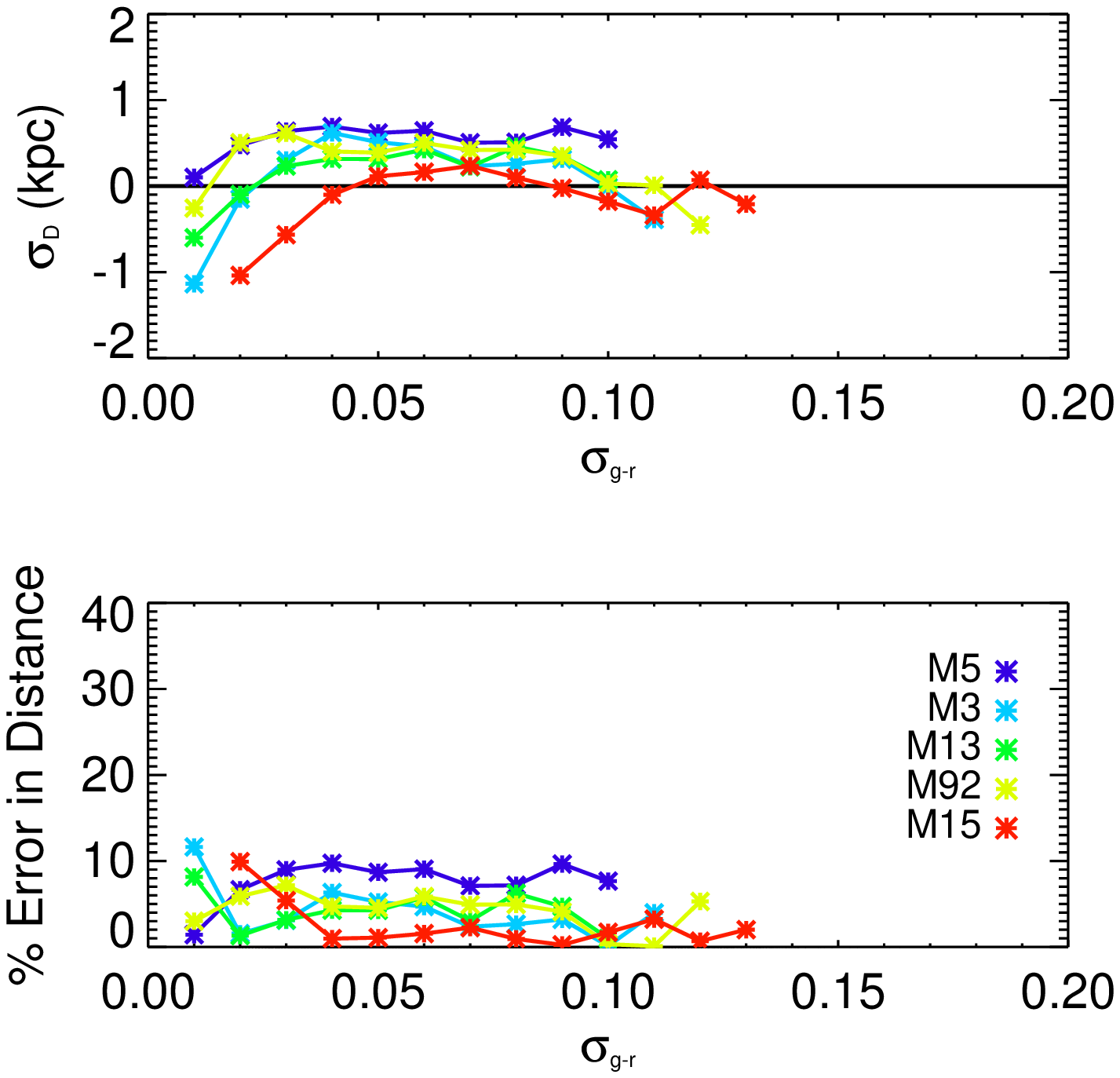} 
\includegraphics[width=1.0\linewidth]{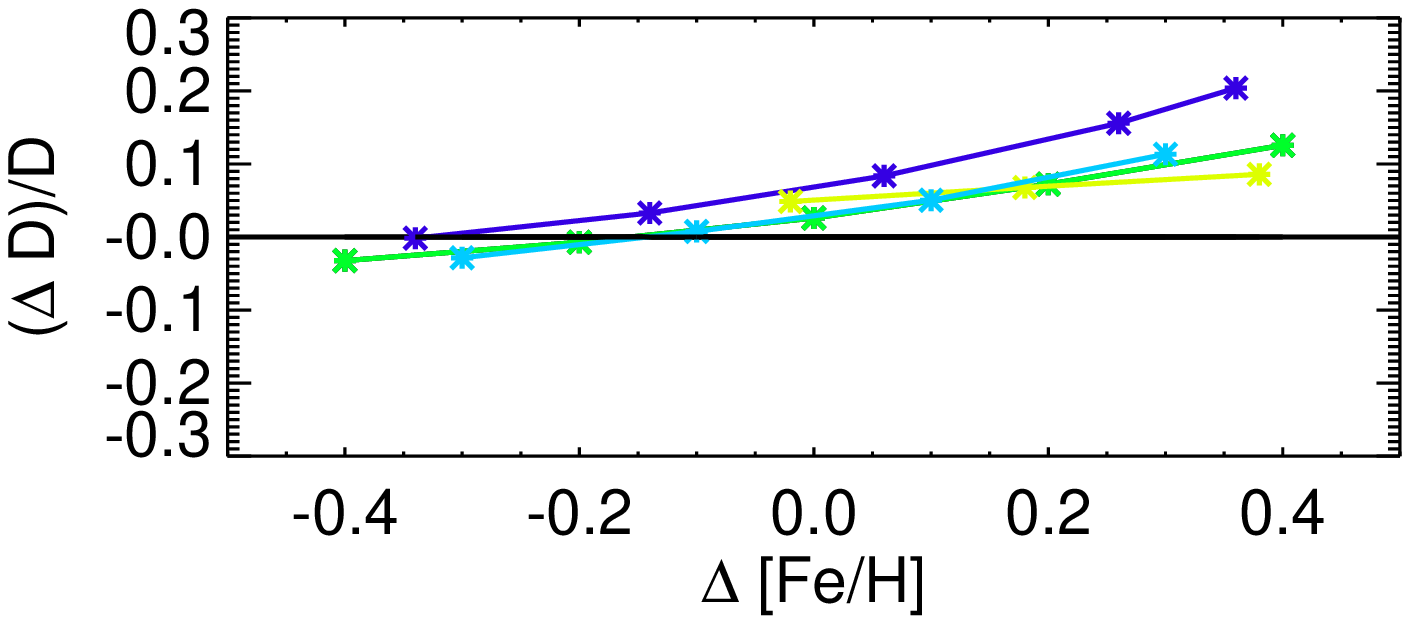} 
 \caption{Errors in distance estimation, as a function of
   uncertainties in photometry and metallicity, are plotted for five
   star clusters (M5, M3, M13, M92 and M15). Top panel: the dispersion, and
   Middle panel: the percentage error in distance due to errors in the
   single-star photometry. The bias in the dispersion at small
   $\sigma_{g-r}$ is due to the mis-identification of sub-giant stars
   near the turn-off as main-sequence stars.  Lower panel: The error
   in distance due to a systematic error in metallicity, determined by
   applying the distance determination method to the star cluster
   photometry while adopting an isochrone with [Fe/H] offset from the
   notional cluster value.  The systematic distance error
   corresponding to 0.2dex uncertainty for [Fe/H] is less than 10\%.
 }
\label{GCs_error} 
\end{center} 
\end{figure}

\begin{enumerate} 
 
\item{Distance error due to photometric errors}

Figure \ref{GCs_error} (top and middle panels) shows the dependence of
error in distance on error in $g-r$ colour, derived from analysis of
the photometry for the five globular clusters. We calibrate errors
based only on the old globular clusters as these are more comparable to WGN
data in age than are the open clusters.  Although the photometric
uncertainties for our field stars are typically less than 3\%, for the
crowded fields of star clusters errors can be larger, allowing
sufficient dynamic range for this test. On average, 80\% of the cluster
stars have $\sigma_{g-r}$ less than 0.1.

The systematic trend in the dispersion of distance errors at small
$\sigma_{g-r}$ is due to two reasons. The dominant effect is the
systematic mis-identification of sub-giant stars near the turn-off as
main-sequence stars below the turn-off, which produces systematically
offset distances. The second, and smaller, of the two effects is the
systematic sensitivity to age near the turn-off, with the method
showing some bias to younger isochrones weighted by a higher
probability from the luminosity function. Not surprisingly,
restricting the isochrone search to the age adopted for the parent
cluster diminished the distance bias.  Of course, a single age cannot
be applied to the mixed population field star WGN data, for which only
a reasonable lower limit of age can be imposed. This distance bias,
though, defines our error limits, and is sufficiently small as not to
limit our later analysis.
  
\item{Distance error due to uncertainty in [Fe/H]} 
 
In our method, we have assumed that the [Fe/H] estimate is accurate,
and hence choose the isochrone with metallicity closest in value to
this measured value. Here we quantify the uncertainty in the distance
introduced by uncertainties in metallicity , which we determined to be
$\sim 0.2$ dex in [Fe/H]. We do this by applying the distance
determination method to the stars in the test clusters while varying
the [Fe/H] of the chosen isochrone around the best value. The bottom
panel of Figure \ref{GCs_error} shows the results in the case of four
of the globular clusters (M15 is not shown here as it has same [Fe/H]
as M92). The WGN and SDSS data,to which our method will be applied,
have [Fe/H] accuracies $\sim 0.2$ dex, and from the figure we determine
that the systematic error on distance due to 0.2 dex uncertainty in
[Fe/H] is within 10\%.

\item{Distance error due to luminosity mis-classification of stars} 

By restricting the isochrone fit only to the main sequence region, we
ensure that no main sequence star is mis-classified as a giant (as
illustrated in Figure \ref{OCs_performance_NGC2420}(b)). But the
sub-giants in the WGN data near turn-off are mostly (more than 95\%)
mis-classified as main sequence stars (as illustrated by the red
points in Figure \ref{GCs_performance_M13}). From
    point (i) in this section we have shown that the there is a
    systematic error due to this mis-classification of sub-giants near
    the turn-off of within 10\%. From Figure \ref{GCs_performance_M13}
    we note that for sub-giants redder than turn off the distance
    errors can blow up to as large as 50\%. 

We also note that in the case of clusters M5 and M92, there is no
upward trend in errors near small $\sigma_{g-r}$, as compared to the
noticeable trend in M3,M13 and M15. This is because of the presence of
fewer giants (less that 0.5\%) in the data of the former two globular
clusters, as compared to the latter in which more than 4.5\% of the
cluster members are giants.

The fraction of a field star sample which will be     sub-giants rather than MSTO stars was a popular topic of     discussion in the 1980's star-count literature. Some of many     papers discussing the issue include \citet{JB85}, \citet{b7},     \citet{F89}, and \citet{GWK}. All are consistent with results that     can now be readily reproduced using the on-line  Besan\c{c}on star     count model as shown in Figure \ref{Besancon_perc}. This confirms that sub-giant contamination in a   colour and magnitude selected survey of thick disk and halo stars with photometric criteria like those of this WGN sample will be     typically at the 5\% to 10\% level, increasing to 13-15\% in some lines of sight. 

Thus we can say that sub-giants, certainly present in the WGN data, will be mis-classified as main sequence turnoff stars, introducing a systematic underestimation of their distance by 20 to 50\%. We expect this contamination from sub-giants to be small (below 10\% per field) and thus not affect our present analysis in a significant way. 

\begin{figure} 
\begin{center} 
\includegraphics[width=1.0\linewidth]{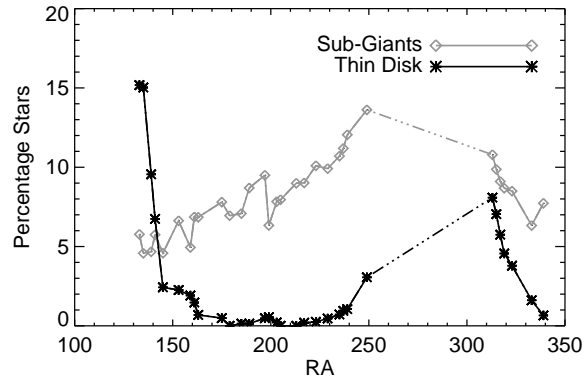} 
\caption{Percentage of sub-giants (Grey Diamonds) and thin disk (Black asterisk) stars in each field as produced by the Besan\c{c}on Model of the Galaxy on application of the WGN selection function.  This gives an estimate of the expected contamination from subgiants and thin disk for a colour and magnitude selected survey of thick disk and halo stars with photometric criteria like those of this WGN sample. The plot confirms that sub-giant contamination in the WGN sample will be typically at the 5\% to 10\% level, increasing to 13-15\% in some lines of sight (discussed in Section 3.2.2.4). It also shows that thin disk contamination is expected to be minimal in the WGN sample (discussed in Section 4.3), ranging between 5 and 15\% in the low latitude fields (i.e near Ra$\sim133^{\circ}$,Ra$\sim249^{\circ}$ and Ra$\sim313^{\circ}$).}
\label{Besancon_perc} 
\end{center} 
\end{figure}

\end{enumerate} 
 
\subsubsection{Distance estimates for the WGN data}
We thus derive distances for stars in the WGN dataset via the
isochrone fitting method described above using $g-r$ and $r-i$
colours. We use a set of 17 isochrones from the Dartmouth Stellar
Evolution Database \citep{b5} for metallicities in the range -2.5 to
0.0 dex (and no alpha-enhancement), such that the choice of isochrones
evenly samples colour-magnitude space. We fit to only the main
sequence of the isochrones and restrict their ages to greater than
10GYrs. The corresponding luminosity functions (age and metallicity
dependent) were adopted from the same database. A $4\sigma$ colour
bracket was required to find matches for lower main sequence stars
with small $\sigma_{g-r}$ where the isochrone sampling is sparse.
Distances could be estimated for 8965 objects out of the 9404 in the
sample. The few hundred stars for which the routine was unable to
determine distances are those which do not fall on the thick disk/halo
CMDs either because they are extreme blue stars ($g-r < 0.2$) or
because they are likely binaries, variables or have mis-calibrated
metallicities. The prior assumed for the age (older than 10GYrs) also
produces a bias against metal-rich stars. The distance distribution
peaks at just above 2 kpc, most stars being within a heliocentric
distance of 6 kpc.

We now consider adding additional photometric information, as a
further test of distance scale systematics.
 
\paragraph{Including infrared photometry - testing systematic distance bias}

\begin{figure*} 
\begin{minipage}{160mm} 
\begin{center} 
\includegraphics[width=1.0\linewidth]{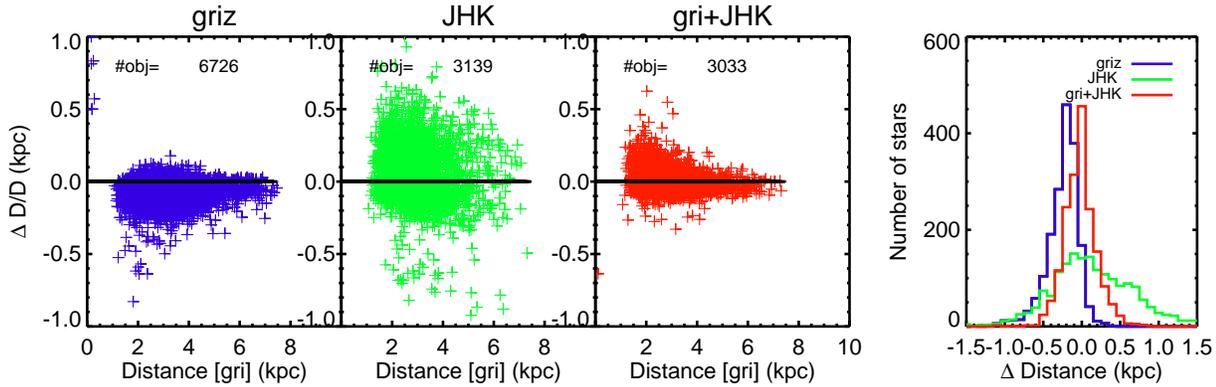} 
\caption{This figure compares isochrone-fit distance estimates using
  optical only, infrared only, and combined optical plus infrared
  photometry.  The first three panels plots the relative error in
  distance estimates from using $griz$ filters (blue), the UKIDSS
  infrared JHK filters (green) and by combining the $gri$ and JHK
  filters (red) with respect to distance estimates using $gri$ filters
  ($D_{gri}$). $D_{griz}$ has a systematic bias, being offset by
  around 0.2 kpc from $D_{gri}$. $D_{JHK}$ shows a mean offset of 0.08
  kpc with respect to $D_{gri}$, though with a large scatter of
  $1\sigma= 0.48$ kpc. This greatly improves in the case of
  $D_{griJHK}$ where the mean offset reduces to -0.03 kpc and a much
  tighter spread of $1\sigma= 0.17$ kpc.  }
\label{ukidss} 
\end{center} 
\end{minipage} 
\end{figure*} 

In order to further check the robustness of our distance estimates,
particularly against systematic bias, we tested the value of
incorporating infrared photometry into the isochrone fitting,
combining J-, H- and K-band data with the $gri$-band data.
 
UKIDSS may be considered as the near-infrared counterpart of the SDSS
Imaging survey \citep{b27}. The Large Area Survey (LAS) covers 4000
sq.~deg. at high Galactic latitudes, with imaging in YJHK filters to a depth of
K=18.4. This coverage includes a LAS equatorial block which overlaps
with most of the sky coverage of the WGN survey.

We cross-matched the WGN dataset with the UKIDSS LAS data. This
provided 3409 matches out of 9404 stars in WGN. We used the JHK
filters in tandem with the $gri$ filters to estimate the distances by
our isochrone fitting method i.e. we now use $x=g-r,r-i,J-H,J-K$ in
Equation 3. The distance determination proceeded as described
above. We compare the new distance estimates, obtained using JHK both
independently and with $gri$, referred to as $D_{JHK}$ and
$D_{griJHK}$, to the distance estimates from just $gri$ ($D_{gri}$) in
Figure \ref{ukidss}.

We see that $D_{JHK}$ shows a mean offset of 0.08 kpc with respect to
$D_{gri}$, though with a large scatter of $1\sigma= 0.48$ kpc. This
greatly improves in the case of $D_{griJHK}$ where the mean offset
reduces to -0.03 kpc and a much tighter spread of $1\sigma= 0.17$ kpc
This shows that there is no bias in the estimated distances using
$gri$ relative to the combined optical-infrared approach. 
In this same figure, we also illustrate the effect of retaining the
$z$ filter in the left most panel. $D_{griz}$ has a systematic bias,
being offset by around 0.2 kpc from $D_{gri}$.

Thus by comparing our distance estimates with those obtained using
tighter constraints (i.e. using $JHK$ along with $gri$), we confirm
that there is no systematic bias in our distances.

\paragraph{Comparing isochrone-fit distances with photometric parallax
distances}

\begin{figure*} 
\begin{minipage}{160mm} 
\begin{center} 
\includegraphics[width=1.0\linewidth]{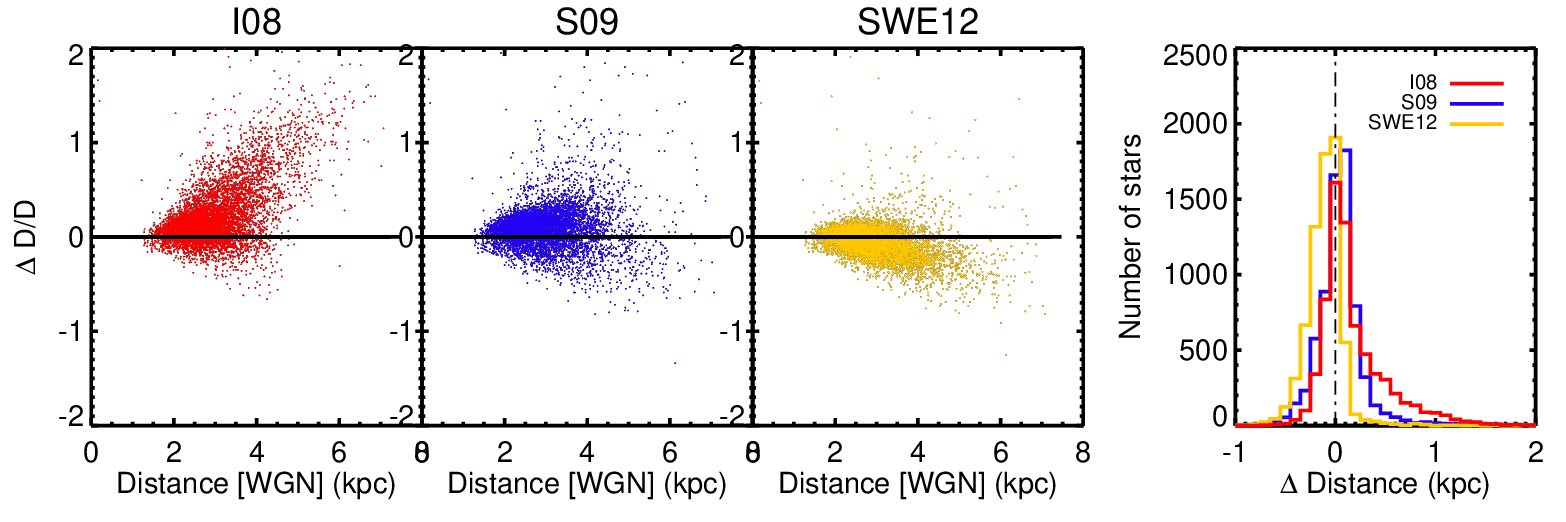} 
\caption{We compare the photometric parallax method of estimating
  distances to the isochrone fitting technique used in this work. In
  the first three panels distance estimates from our isochrone fitting
  method are compared to distances from photometric parallaxes, using
  the recipes prescribed in \citet{b13} (red), \citet{b22} (blue), and
  \citet{b21} (orange) in terms of relative distance error. I08
  systematically overestimates the distances with an average offset of
  $\sim 0.09$ kpc, showing larger deviations at larger distances. The
  match improves in the case of S09 with a mean offset of $\sim 0.08$
  kpc, while the SWE12 method matches the distances computed in this
  work most closely. The mean offset with respect to SWE12 is $\sim
  -0.04$ kpc, with a spread of $\sim 0.12$ kpc.}
\label{PP} 
\end{center} 
\end{minipage} 
\end{figure*}

We also compared our distances to those obtained by applying the
method of photometric parallax used by \citet{b13} (hereafter I08). In
this method, an empirical relationship between absolute magnitude and
colour is first adopted, which is independent of both age and
metallicity and only describes the $shape$ of the colour-magnitude
sequence below turn-off. This zeroth-order relationship is usually
adopted by finding the best-fit function to the CMD of a globular
cluster of known abundance and whose distance is known fairly
accurately by means of some other reliable method of distance
measurement. Then correction terms are introduced which shift this
relation to represent the true age and metallicity.
 
I08 perform an empirical fit simultaneously to the photometry of stars
of five globular clusters from SDSS to obtain the basic photometric
parallax relation. They then use the M13 globular cluster to determine
the correction term due to the effect of age on the CMD. This age
correction is calculated under the constraint of a fixed value of
metallicity - which in their case is the mean metallicity of M13
([Fe/H] = -1.54), which is close to the mean metallicity of the
Galactic halo. Strictly speaking, this will not hold for thick disk
stars whose mean metallicity, at least in local samples, differs by the
order of +1.0 dex from that of halo stars, even if the two populations
have approximately the same age, within an uncertainty of 2 GYr. I08
adopt this correction on the basis that this will produce an error of
less than 0.2 mag for disk-like stars (which corresponds to an error
in distance estimate of 1 kpc in our case).  \citet{b22} (hereafter
S09) propose an improved correction which takes into account that the
morphology of the colour-magnitude relation near turn-off depends
significantly on both age and metallicity. More recently \citet{b21}
(hereafter SWE12) modified this to construct a global turn-off
correction applicable to a mixture of disk and halo stars.  All the
above methods were applicable in the colour region of $0.3 < g-i <
0.6$. This colour restriction eliminates close to 2000 stars from the
WGN dataset, the majority of these being thick disk turn-off stars
with $g-r \ge 0.6$.

Selecting only those stars from the WGN dataset which fall in the
photometric calibration $g-i$ colour range, the distances computed
here using our isochrone matching technique may be compared to the
values obtained by applying the methods of I08,S09 and SWE12. The
comparison is shown in Figure \ref{PP}. I08 systematically
overestimates the distances to the stars, with larger deviations at
larger distances. The I08 values are on average offset from distances
from this work by $\sim 0.09$ kpc. The match improves in case of S09
with a mean offset of $\sim 0.08$ kpc, while the SWE12 method match
the distances computed in this work most closely. The mean offset with
respect to SWE12 is $\sim -0.04$ kpc, with a spread of $\sim 0.12$
kpc.
 
This provides a final consistency check on our distance estimates.   

The method of photometric parallax works well in the case of faint
main-sequence stars and single stellar population, but is limited by
the availability of calibration stars over a large range of stellar
mass and metallicity. In the case of a mixture of populations, the age
correction in the photometric parallax method requires an assumption
about the fraction of stars from different populations which S09, for
example, determine based on a kinematic and chemical segregation of
stars. Our method based on stellar evolution models avoids any such
assumption and is more robust in accounting for the age and
metallicity effects for turn-off stars.

We thus have estimated distances to 8965 stars in
    the WGN dataset, with 90\% of them having uncertainty within
    10\%. 

\begin{figure*} 
\begin{minipage}{180mm} 
\begin{center} 
\includegraphics[width=0.6\linewidth]{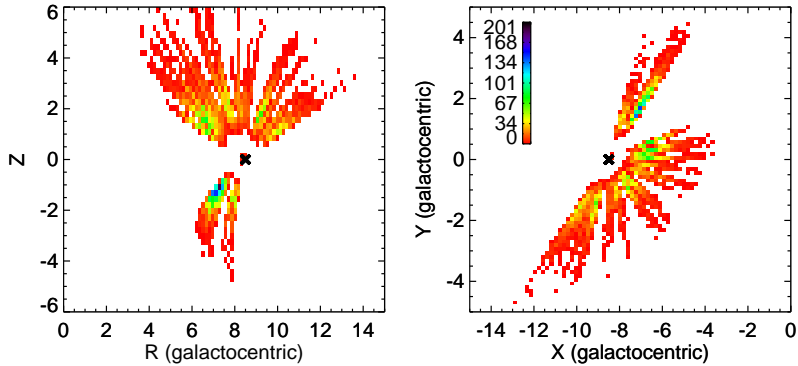} 
\caption{The spatial extent of the WGN survey data in XYZR galactic
  coordinates. In both plots, the black cross shows the position of
  the Sun. Colours indicate the number of stars in each spatial bin,
  colour coded as shown.}
\label{RZXY} 
\end{center} 
\end{minipage} 
\end{figure*}

Figure \ref{RZXY} shows the spatial coverage of the data. The data
covers a range of Galactocentric radii (projected distance on the
plane) between 4 and 12  kpc. There is also coverage above and below
the Galactic plane, extending up to a height of 6  kpc. This provides
the scope to investigate radial and vertical gradients in the
abundance and kinematic parameters, as well as to investigate spatial
distributions in a coherent way.  The coverage of the data on the
Galactic plane, shown by the X vs Y plot, shows the potential of the
data to investigate symmetries in the Galaxy - about the plane, about
the cardinal direction l$\sim 270^{\circ}$ and about the centre of the
Galaxy.

\subsection{Proper Motions and Global Space Velocities} 

Proper motions (along RA and Dec) for the WGN stars are available from
SDSS DR7. Using these along with the distance estimates obtained as
described in the previous section and the measured radial velocities,
we can obtain right-handed Galactic U,V,W velocities using the
conversion outlined in \citet{b14}. These are then converted to
cylindrical coordinates $V_R, V_\phi, V_Z$ by simply rotating the
coordinate system about the W axis.
 
In the cylindrical coordinate system, the contribution of the three
velocity components to the line of sight (LOS) velocity at any point
in space specified by the Galactic longitude $l$, Galactic latitude $b$
and distance from the sun D are given as
\begin{equation} 
V_{LOS}= A_R V_R + A_{\phi} V_{\phi} + A_Z V_Z  
\label{eq_vLOS} 
\end{equation} 
where 
\begin{eqnarray} 
A_R &=& \mathrm{sin}(\alpha)\mathrm{cos}(b)\\ 
\nonumber A_{\phi}&=&\mathrm{cos}(\alpha)\mathrm{cos}(b) \\ 
\nonumber A_Z&=& \mathrm{sin}(b) 
\label{eq_componenets} 
\end{eqnarray} 
and where $\alpha$ is the angle between $V_\phi$ and the
line of sight direction which depends on l,b and D and is given as follows:
\begin{eqnarray}
\alpha=\pi - l - \theta_{GC}\\
\nonumber \mathrm{where}~~ \theta_{GC}&=&\mathrm{tan}^{-1} \frac{D
  \mathrm{cos(b)}\mathrm{sin(l)}}{R_{0} -
  D\mathrm{cos(b)}\mathrm{cos(l)} } 
\end{eqnarray}
and $R_0$ is assumed to be 8.5 kpc \citep{bref}. 

Here, $V_{LOS}$ is the line of sight velocity
component in the rest frame of the Galaxy which is related to the
measured heliocentric radial velocity $V_{hel}$ by  
\begin{equation} 
V_{LOS}=V_{hel}+V_{corr}+V_{LSRcorr} 
\end{equation} 
where $V_{corr}$ corrects for the peculiar motion of the sun
[we adopt $(U_{\odot},V_{\odot},W_{\odot}) = (11.1,12.2,7.2)$ km/s \citep{b19}]
with respect to the Local Standard of Rest  
\begin{equation} 
V_{corr}=U_{\odot}\mathrm{cos}(l)\mathrm{cos}(b) +V_{\odot}\mathrm{sin}(l)\mathrm{cos}(b)+ W_{\odot}\mathrm{sin}(b)  
\end{equation} 
 and $V_{LSRcorr}$ converts the measured quantity into the rest frame of the Galaxy:
\begin{equation} 
V_{LSRcorr}=V_{LSR}\mathrm{sin}(l)\mathrm{cos}(b)  
\end{equation} 
$V_{LSR}$ has been taken to be 220 km/s \citep{bref} .
 
Given the spatial distribution of stars in our dataset, the relative
contributions of the three Galactic Space velocity components to the
measured line of sight velocity varies widely across the fields.  The
contribution of $V_\phi$ goes to zero in the direction towards the
Galactic Centre, while the $V_R$ contribution goes to zero
perpendicular to this direction ($l \sim 90^{\circ}$ and
270$^{\circ}$). This is reflected in the variation of uncertainties in
the velocities of each of these components along Galactic
longitude. While $V_R$ is most accurately determined towards the
Galactic centre from $V_{LOS}$ alone, $V_\phi$ is more reliably
determined in the perpendicular direction.

\begin{figure*} 
\includegraphics[width=1.0\linewidth]{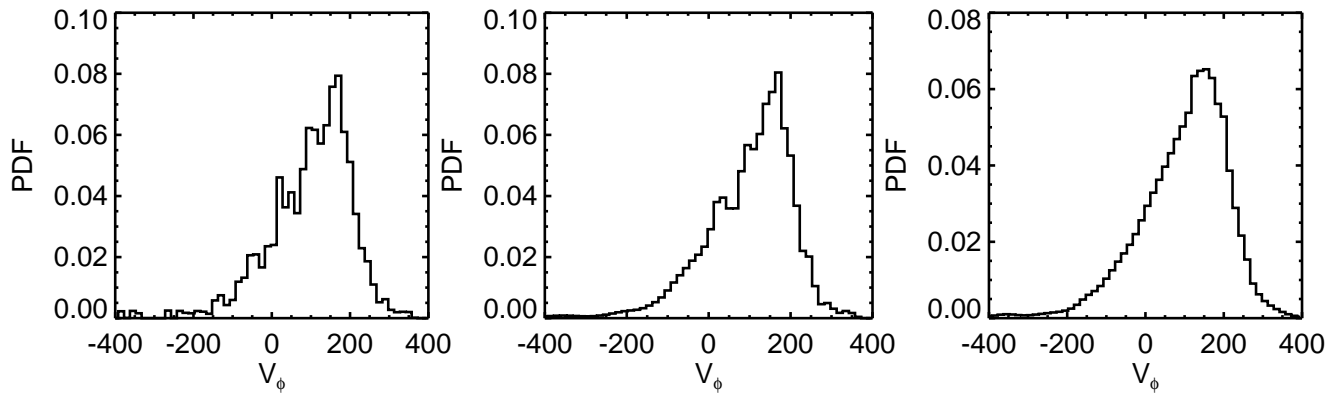} 
\caption{Space motion error analysis for fields in the direction
  $l=270^{\circ}$. Left Panel: $V_\phi$ distribution derived using
  radial velocity, proper motions and distance, while assuming that
  there are zero errors on distance and proper motions. Middle panel:
  $V_\phi$ distribution when convolved with the contribution of error
  from distances alone (i.e. setting
  $\sigma_{\mu_\alpha},\sigma_{\mu_\delta}=0$). Distance errors are
  less than 10\%, thus ensuring that 80\% stars have less than 10\%
  error on $V_\phi$. Right panel: $V_\phi$ distribution when
  convolved with the errors from proper motions alone
  (i.e. $\sigma_D=0$) which are typically 3mas/yr resulting in 85\%
  stars having greater than 20\% error on $V_\phi$. Evidently proper
  motion uncertainties dominate the error budget.}
\label{erroranal} 
\end{figure*}

In this work, we are particularly interested in the global behaviour
of the $V_{\phi}$ parameter, as this is a primary thick disk-halo
discriminator. Since the sensitivity of the line of sight velocity to
$V_{\phi}$ varies considerably across our fields, large errors in
proper motion can significantly distort the true $V_{\phi}$
distribution. Figure \ref{erroranal} analyses the contribution of
errors from the two different sources, namely from distance and proper
motion errors, to the $V_{\phi}$ distribution. Considered in this
figure are only fields centred on the cardinal direction $l = 270
^{\circ}$ , which makes the observed radial velocity more sensitive to
$V_{\phi}$ as compared to $V_R$. The leftmost figure shows the
distribution of the computed $V_{\phi}$ if both distance and proper
motion measurements are considered to be accurate. The middle and
rightmost plots show how the Probability Distribution Function (PDF)
is altered if there is a contribution of error from only distances and
only proper motions respectively. The typical errors in proper motion
are around 3mas/yr while typical errors in distance are 0.2 kpc. When
considering errors only from proper motions (i.e. setting
$\sigma_D=0$) only 15\% of the stars have $\sigma_{V_\phi}$ within
10\%. The typical errors on $V_{\phi}$ are $\sim 13$\%. Almost 40\% of
the stars have errors on $V_{\phi}$ greater than 20\% thus smearing
out the distribution on the rightmost panel of Figure
\ref{erroranal}. In contrast, when considering errors only from
distances (setting $\sigma_{\mu_\alpha},\sigma_{\mu_\delta}=0$) 80\%
of the stars have $\sigma_{V_\phi}$ better than 10\% with typical
errors of 3\%.

This clearly illustrates that proper motions are the chief
contributors to errors in the velocity components, a factor we
consider in the analysis below.

\section{Analysis} 

We now proceed to analyse the WGN survey data, considering in turn the
validity and numerical values of single parametric descriptions of the
thick disk, for kinematics and spatial scale length. We then consider,
and discover, spatially localised substructure in the survey.

\subsection{Kinematic and Number-Density Modelling in Observable Space} 

When analysing a set of distances, abundances and kinematics for a mix
of stellar populations one approach is to minimise error propagation
by looking for models in physical space that most closely reproduce
probability distribution functions in observable space.  In our
sample, the radial velocities are of higher internal precision than
are the other contributions to space motions (cf. Figure
\ref{erroranal}), so we proceed accordingly.
 
We model the data in the one dimensional space of line of sight
velocity $V_{LOS}$. The data are assumed to be completely described by
a two-component model consisting of the thick disk (denoted as TD) and
the halo, using the discussion associated with the selection function
earlier to justify exclusion of thin disk populations. Initially, the three
cylindrical velocity components i.e. $V_R,V_{\phi},V_Z$ are assumed to
be distributed as Gaussians for the thick disk and halo independently.
 
A maximum likelihood method is employed to determine the best model
given the data. The likelihood of a particular model $\Phi$ given n
data points $X=x_1,x_2...,x_n$ is then given by
\begin{equation} 
 L(\Phi|X)=\prod_1^n P(x_i|\Phi) 
\end{equation} 
where  
\begin{eqnarray} 
\nonumber P(x_i|\Phi)&=& f P(x_i|\Phi^{TD})+(1-f) P(x_i|\Phi^{halo})\\ 
\nonumber &=& f P(\Phi^{TD}(V_{LOS,i}|V_{{LOS}}))\\
\nonumber & &  +(1-f) P(\Phi^{halo}(V_{LOS,i}|V_{{LOS}}))
\end{eqnarray} 
with $f$ being the normalisation of the thick disk component with respect to
the total number of stars.
 
Since we do not know the shape of the distribution of $V_{LOS}$ we can
recast this in terms of $V_R,V_{\phi},V_Z$ which have (assumed) Gaussian
distributions, and then marginalise over the unknown parameters. This
is done using the method outlined in \citet{b20} in their Appendix A.

The probability of observing a star with a line of sight velocity
between $V_{LOS}$ and $dV_{LOS}$ is given by
\begin{eqnarray} 
\nonumber
\phi(V_{LOS})dV_{LOS}=\int_{-\infty}^{\infty}\int_{-\infty}^{\infty}\phi_R(V_R)\phi_\phi(V_\phi)\phi_Z(V_Z)
dV_R dV_{\phi} dV_Z 
\label{eq_sirko} 
\end{eqnarray} 
with  
\begin{equation} 
\phi_x(V_x)dV_x = \sqrt{\frac{1}{2\pi\sigma_x^2}}\exp({\frac{-(v_x-\mu_x)^2}{2\sigma_x^2}}). 
\label{eq_gauss} 
\end{equation} 
 
Using the form of $V_{LOS}$ given by equation \ref{eq_vLOS},
equation \ref{eq_sirko} can be analytically solved to obtain the
following simple solution
\begin{equation} 
\phi(V_{LOS})=\frac{1}{\sqrt{2\pi\Sigma}}\exp[-\frac{\lambda^2}{2\Sigma}] 
\end{equation} 
where 
\begin{eqnarray} 
\nonumber \Sigma=\sigma_R^2A_R^2+\sigma_{\phi}^2A_{\phi}^2+\sigma_Z^2A_Z^2,\\ 
\nonumber \lambda=-V_{LOS}+\mu_RA_R+\mu_{\phi}A_{\phi}+\mu_ZA_Z 
\end{eqnarray} 
with $A_R,A_{\phi},A_Z$ being defined in Equation 6. 

The best relative normalisation $f$ is obtained independently for each
field for any given configuration of all the other four free
parameters.

\subsection{Maximum likelihood determination of the thick disk
  kinematic parameters}
 
In this analysis we seek first to find a maximum likelihood
description of the velocity components of the thick disk. The free
parameters in the model are $\mu_{V_\phi}$ (henceforth written simply
as $V_\phi$) and $\sigma_{V_R},\sigma_{V_\phi},\sigma_{V_Z}$ of the
thick disk, and $f$, the relative number of thick disk stars to halo
stars for each field. All kinematic parameters describing the halo,
together with the mean $Z$ and $R$ velocities of the thick disk, are
held as fixed parameters in the model. The mean $Z$ and $R$ velocities
of the thick disk and mean motions of the halo are set to zero,
assuming equilibrium and that there is no net rotation in halo. All
those $\sim 8900$ stars with good photometry, $\sigma_{g-r} \le 0.1$
and weight error less than 50\% were included in the modelling.  To
ensure kinematic parameters appropriate to SDSS survey results
covering comparable volumes to those investigated here, we adopt the
values of the velocity dispersions of the inner halo determined by
\citet{b4}.  These values are summarised in Table \ref{Table_par}.

\begin{table*}
 \begin{minipage}{140mm}
\begin{center}
  \caption{Kinematic parameters adopted and determined in the Galactic model} 
  \begin{tabular}{@{}lccl@{}} 
 \hline 
   Parameter &  Value fixed (km/s) & Value determined (km/s) & Reference\\ 
 \hline 
Halo \\ 
${V_R}, {V_\phi}, {V_Z}$ & 0, 0, 0     & & assumed \\ 
$\sigma_{V_R}$           & 150 $\pm$ 2 & & \citet{b4}\\
$\sigma_{V_\phi}$        & 95 $\pm$ 2  & & \citet{b4}\\
$\sigma_{V_Z}$           & 85 $\pm$ 1  & & \citet{b4}\\ 
 \hline 
Thick Disk \\ 
${V_R}$, ${V_Z}$ & 0, 0 & & assumed \\
$V_{\phi}, \sigma_{V_{\phi}}$ & & 172 $\pm$ 2, 49 $\pm$ 2 & this study \\
 $\sigma_{V_R}$               & & 51 $\pm$ 3              & this study \\
$\sigma_{V_Z}$                & & 40 $\pm$ 3              & this study \\ 
\hline 
\label{Table_par}

\end{tabular}
\end{center} 
\end{minipage}
\end{table*}

\begin{figure*} 
\begin{minipage}{180mm} 
\begin{center} 
\includegraphics[width=0.6\linewidth]{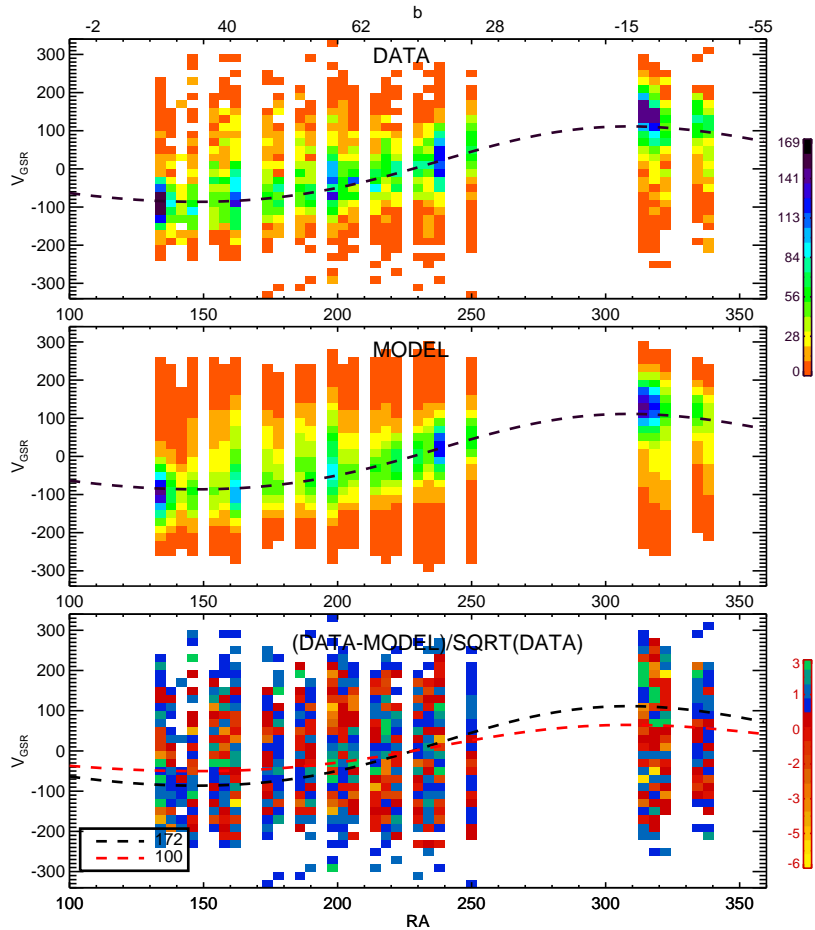} 
\caption{Galactocentric velocity vs RA is plotted for the weighted
  data (top panel) and for the model with our best fit thick disk
  rotational parameters of $V_{\phi}, \sigma_{V_{\phi}} = 172,49$km/s
  and $\sigma_{V_R},\sigma_{V_Z} =51,40$km/s (middle panel). For these
  two panels, colours indicate the number of stars in each bin, as
  indicated. The dashed lines correspond to a rotational velocity of
  172 km/s, the determined best fit mean, at 2 kpc. Both data and
  model plots have been normalised to have 1000 stars in each field in
  order to bring out features more clearly. All the velocity
  parameters of the halo are fixed in the model. (see Table
  \ref{Table_par}). The relative normalisation between thick disk and
  halo for each field is allowed to be a free parameter. Lower panel:
  this shows the significance of the difference between the model and
  the data. In this panel, blue to green denotes positive deviation
  ranging from $0$ to $+3\sigma$ while red to orange denotes $0$ to
  $-6\sigma$.  The dashed lines in the lower panel correspond to a
  rotational velocity of 172 km/s and 100 km/s at a distance of 2 kpc,
  and are illustrative. Galactic latitude is also indicated at the top
  of upper panel.}
\label{result1} 
\end{center} 
\end{minipage} 
\end{figure*} 

\begin{figure*}
\begin{minipage}{180mm}
\begin{flushright} 
\includegraphics[width=0.8\textwidth]{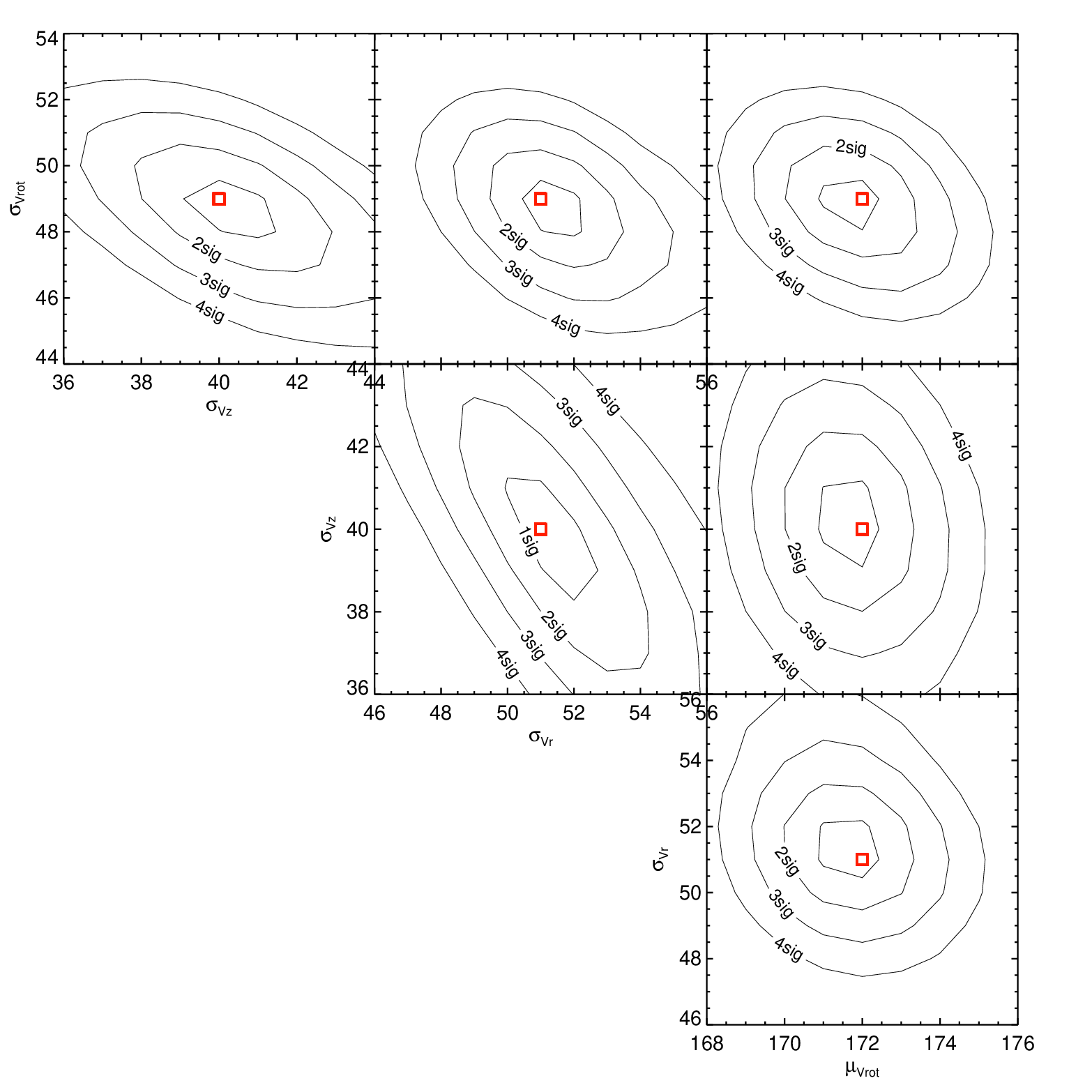} 
\caption{The likelihood contours of the four thick disk kinematic
 parameters determined in the maximum likelihood fitting of the model -
  $V_{\phi},\sigma_{\phi},\sigma_{R},\sigma_{Z}$. The red box in
  each case marks the value for which the likelihood is maximised.}
\label{contour}
\end{flushright} 
\end{minipage} 
\end{figure*} 
 
Figure \ref{result1} shows the Galactocentric velocity distribution
for the data and the corresponding distribution produced by the best
fit model resulting from our maximum likelihood analysis, where the
best fit values were found by scanning through parameter space with a
step size of $\mu,\sigma$= 1,1 km/s. In this plot, the number of stars
is normalised to be the same in each field in order to pick out
features easily by eye. The best fit parameters recovered by the
Maximum Likelihood modelling were ${V_\phi},\sigma_{V_\phi}$ = 172, 49
(km/s) and $\sigma_{V_R},\sigma_{V_Z}$ = 51, 40 (km/s).  The
rotational velocity is only marginally lower than a canonical thick
disk rotation ($\sim 180$ km/s, \citet{b4}). The contour plots
localising maxima and indicating error bounds and parameter
correlations for the various parameters are shown in Figure
\ref{contour}. These figures show that this simple kinematic model is
a good global description of the data, while our best-fit kinematic
parameters are in good agreement with earlier determinations
(e.g. \citet{b4}, \citet{Chiba}).
 
The significance of the difference between model and data (i.e
difference between model and data divided by the Poisson error in
number counts in each pixel) is plotted in the bottom-most panel of
Figure \ref{result1}. A significant mismatch, which we define as
several adjacent pixels each of which is more than $2\sigma$ from the
model with the same sign, is evident only near RA $\sim
313^{\circ}$. Here many stars seem to have higher line of sight
velocities than predicted by the model.

The region of mismatch covers $312^{\circ} \le RA \le 318^{\circ}$.
It spans low, southern latitudes $-30^{\circ} \le b \le -26^{\circ}$
and lies in the inner Galaxy $48^{\circ} \le l \le 50^{\circ}$.  It
has radial velocities $\sim$20km/s higher than do the typical thick
disk stars in these lines of sight. Hereafter, we will refer to this
region as R313.

\subsection{Sub-structure in the Galactic thick disc?}
Is the mismatch seen between the global best-fit kinematic model and
the data due to substructure in the thick disk, or is it just an
artefact of incorrect extinction or systematic errors in distance? Are
the stars with high radial velocity with respect to the thick disk just
high contamination from thin disk stars at these low latitudes? We
analyse the possibilities in this section.

R313, being at the lower Galactic latitude range of the WGN fields
that looks into the inner parts of the Galaxy, is likely to have a
larger contamination of thin disk stars. The high rotational velocity
of the thin disk will show itself with radial velocities higher than
that of thick disk by similar amount as seen in this direction. To
investigate whether an inadequate model of thin disk stars contributes
to the mismatch, one may consider comparing R313 with the
correspondingly low Galactic latitude WGN field near RA$\sim
133^{\circ}$ ($\ell=228^{\circ}, b=+27^{\circ}$).  We found,
especially by comparing the metallicity distributions in these two
directions, that there is no significant contribution from a thin disk
population. By looking at the number of stars in
    each field as given by the  Besan\c{c}on Model of the Galaxy
    \citep{b18} on applying the WGN colour-magnitude cuts, we verify
    that the number of thin disk stars that pass the selection
    function is minimal , the contamination being close to 0 in most
    of the high latitude fields and between 5 and 15\% in all the
    lowest latitude fields, as shown in Figure \ref{Besancon_perc}. The thin disk kinematic signal, if
    present, may be unidentifiable towards the Galactic centre, near
    fields RA$\sim249^{\circ}$, but RA$\sim133^{\circ}$ being
    similarly inclined to the centre-anti-centre line as
    RA$\sim313^{\circ}$, we would expect the thin disk stars to
    equally show themselves in the line of sight velocity
    distributions in these fields as well, but this is not the
    case. The data match our kinematic model well near RA $\sim
    133^{\circ}$. Hence we conclude that a misrepresentation of the
high-latitude symmetric thin disk cannot be the cause of the excess
numbers of stars with high line of sight velocities. For similar
reasons, the fact that we may have ignored any vertical velocity
gradient in the kinematic model for the thick disk is an unlikely
cause of the mismatch.

Might incorrect interstellar extinction be at fault?  If one
underestimates the extinction one overestimates the distance, which
can make stars in this region appear to be moving faster. The
estimated distance of stars in R313 is on average between 2 to 3
 kpc. However, even a 25\% distance error increases the $V_{GSR}$ at
$RA \sim 313^{\circ}$ only by 5 km/s for a rotational velocity of 180
km/s. Therefore, overestimated distances due to incorrect extinction
cannot produce as significant a shift in $V_{GSR}$ as is being
observed here. This would also not explain the excess star counts discussed in the next section.
 
The above checks confirm that the velocity mismatch seen in region
R313 is real and due to some substructure.

\subsection{Modelling number counts - the thick disk scale length}

\begin{figure*} 
\begin{minipage}{180mm} 
\begin{center} 
\includegraphics[width=0.9\linewidth]{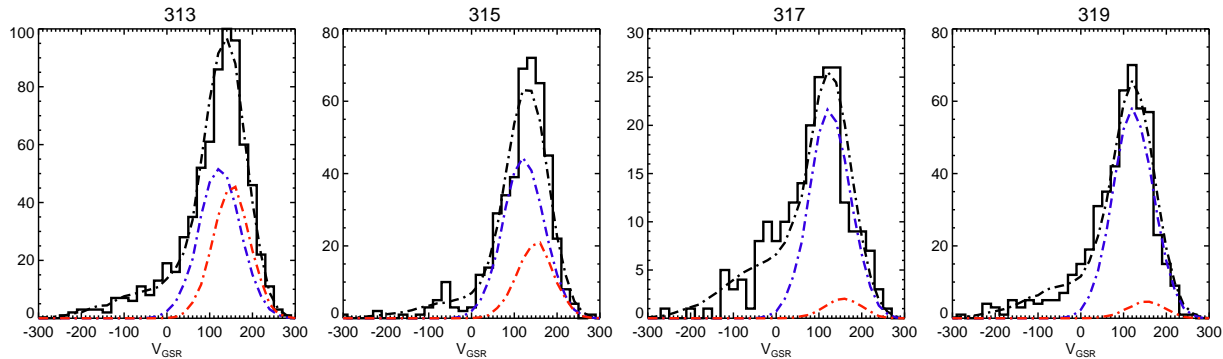} 
\caption{The panels show line of sight velocity for four fields near
  R313 where the mismatch in velocities between data and model is
  apparent. Data in these fields are simultaneously fit by three
  components - thick disk, halo and a third component representing the
  overdensity. Halo kinematics are held fixed as before (Table
  \ref{Table_par}). Thick disk kinematics are adopted from the fit
  result in Section 4.2 i.e. $V_{\phi},
  \sigma_{V_{\phi}},\sigma_{V_R},\sigma_{V_{Z}}$ = 172,49,51,40 km/s
  with mean $V_R,V_Z$ fixed to zero. The overdensity is assumed to be
  thick disk like, albeit with a different rotation velocity. All
  parameters for the overdensity are fixed to thick disk values,
  except$V_{\phi}, \sigma_{V_{\phi}}$ which are fit by maximum
  likelihood. The best fit value obtained was 206,36 km/s. The
  histograms show the unweighted data, the blue curve show the thick
  disk model, red curve shows the overdensity model and the black
  curve shows the total model in each case. The RA of the field is
  indicated above each panel. The contribution of this extra component is as
  high as 35\% and 30\% in the left two panels while it drops to 5\%
  in the last two.}
\label{OVER_FIT} 
\end{center} 
\end{minipage} 
\end{figure*}

We look at \textit{how many} stars in the R313 fields contribute to
the velocity mismatch. For this, we first assume that the substructure
is disk-like with similar $V_R$ and $V_Z$ distributions as the thick
disk, but a higher mean $V_\phi$. We fix the halo kinematics as before
and fix the thick disk kinematics to the values determined from the
global fit in Section 4.2. We isolate four WGN fields that cover
region R313 and fit for $V_{\phi}$ and $\sigma_{\phi}$ of the
substructure. We obtained a maximum likelihood fit of
$V_{\phi},\sigma_{\phi}$ =206,36 km/s. As shown in Figure
\ref{OVER_FIT} the substructure has a 35\% and 30\% contribution to
the fields centred on RA 313$^\circ$ and 315$^\circ$ respectively
while the contribution falls sharply to 5\% in the fields centred on
RA 317$^\circ$ and 319$^\circ$. We should expect these stars to show
up as an excess in number count distributions of the thick disk.

\begin{figure} 
\begin{center} 
\includegraphics[width=0.9\linewidth]{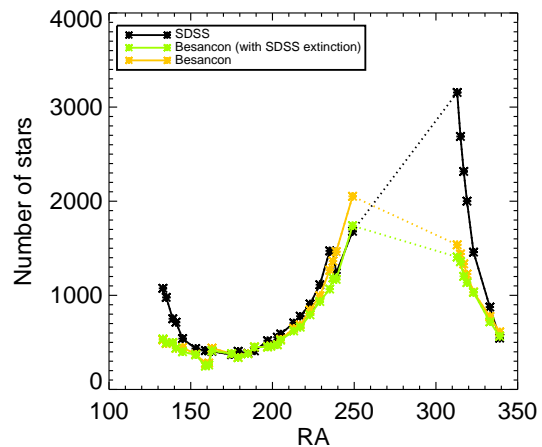} 
\caption{The surface number density distribution of stars per WGN
  field is compared between SDSS DR7 and the  Besan\c{c}on Model of the
  Galaxy for a magnitude range of $16 \le g \le 17.5$ and same colour
  cuts as the WGN selection function. black : SDSS DR7 data ; orange :
   Besan\c{c}on Model values ; green :  Besan\c{c}on model number counts
  corrected with the corresponding extinction for each field from SDSS
  DR7. SDSS data shows an excess of stars in comparison to the
   Besan\c{c}on model in two regions - RA less than 140$^\circ$ and RA in
  the range 313$^\circ$ to 330$^\circ$, which figure \ref{trend}
  shows is the region sensitive to thick disk structural parameters.}
\label{numbercounts} 
\end{center} 
\end{figure}

We therefore go on to analyse the number of stars per field present in
the SDSS DR7 data adopting the same colour-magnitude cuts as the WGN
selection function, and compare this to the number count predictions
given by the  Besan\c{c}on Model of the Galaxy. This comparison
is shown in Figure \ref{numbercounts}, where significant differences
between the model and the data are evident in the RA ranges which
correspond to the lowest Galactic latitudes (RA$\sim$133$^\circ$ and
RA$\sim$313$^\circ$). The default extinction in the  Besan\c{c}on model is
smaller than in SDSS data, so we also show model predictions adopting
the SDSS DR7 mean extinction in each field, confirming a real
difference between model and data.

\begin{figure} 
\begin{center} 
\includegraphics[width=0.8\linewidth]{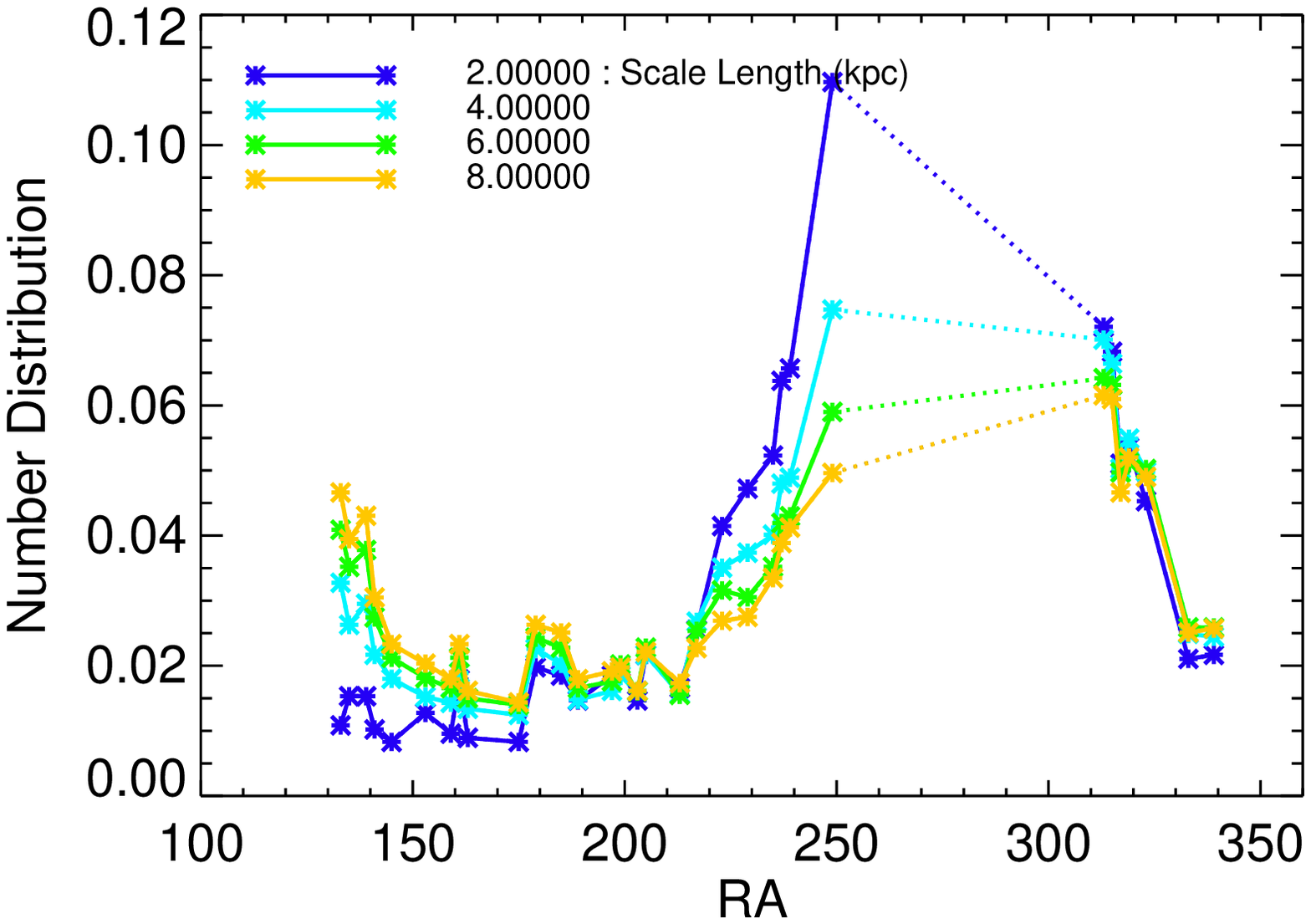}
\includegraphics[width=0.8\linewidth]{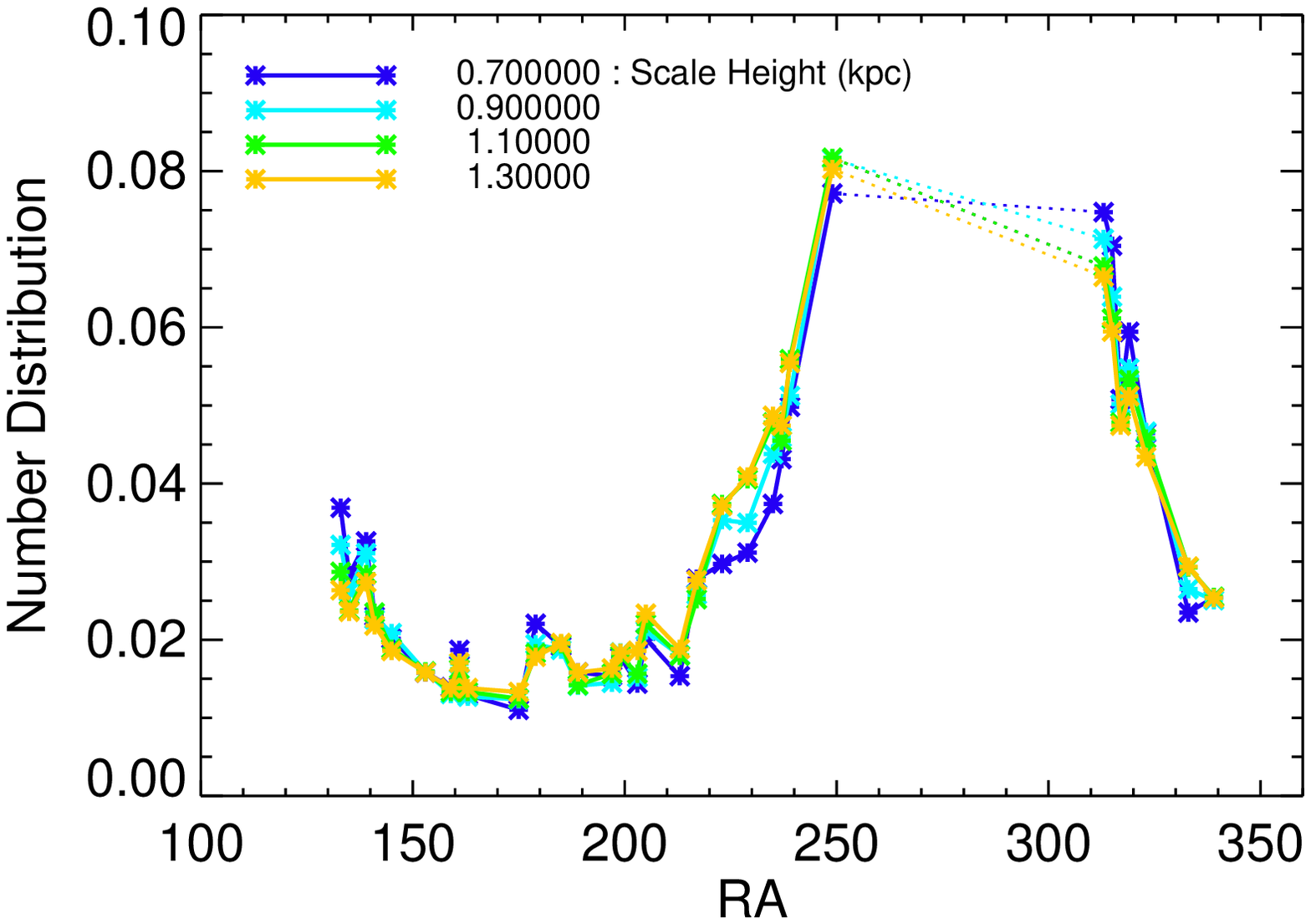} 
\includegraphics[width=0.9\linewidth]{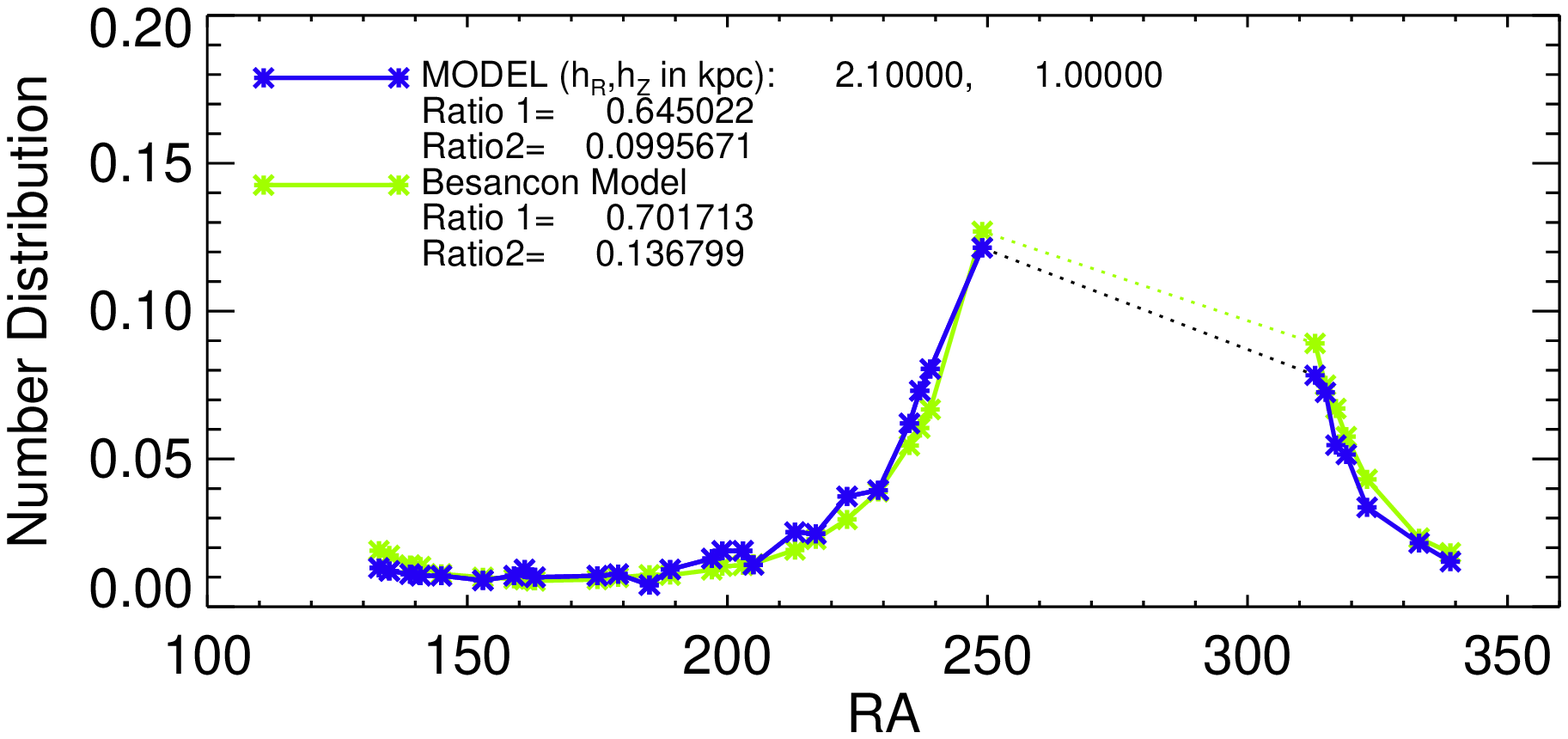}
\caption{A simple illustration of the sensitivity of star counts in
  the WGN fields to the structural parameters of the thick disk.  The
  figure shows number count distributions produced by a simple density
  model for thick disk-halo : a double exponential disk parametrised
  by scale height $h_Z$ and scale length $h_R$ for the thick disk and
  a flattened power-law with flattening $q=0.6$ and index $\alpha=2.3$
  for the halo. Distributions are obtained for the WGN lines of sight
  for stars in a 6 kpc region centred on the Sun. The global ratio of
  halo to thick disk stars in the Galaxy is held fixed. Top: $h_Z$
  fixed at 0.9 kpc. $h_R$ varied from 2 to 8 kpc in steps of
  2 kpc. Middle: $h_R$ fixed at 4 kpc. $h_z$ varied from 0.7 to 1.3 kpc
  in steps of 0.2 kpc. We can see that increasing $h_R$ or decreasing
  $h_Z$ significantly changes the number counts near
  RA$\sim$135$^\circ$ and RA$\sim$315$^\circ$, which correspond to the
  lowest Galactic latitudes probed. Bottom: Comparison of the simple
  parametric star count galaxy model with star count distributions in
  the WGN fields predicted by the  Besan\c{c}on model. The values of the
  two figures of merit defined in the text, Ratio1 and Ratio2, are
  given for the distributions for quantitative comparison. The model
  with $h_R,h_Z$ = 2.0,0.8  kpc (blue) reproduces the number
  distribution of the  Besan\c{c}on model (green) adequately, showing these
  are equivalent exponential scale lengths to the model for present
  comparative purposes.}
\label{trend} 
\end{center} 
\end{figure}

In order to understand the reasons for this difference between model
and data, we describe here a simple toy-model analysis. We generated
the star count density distribution of the thick disk and halo using
respectively a double exponential disk parametrised by scale height
$h_Z$ and scale length $h_R$ to describe the density distribution of
the thick disk and a flattened power-law with flattening $q=0.6$ and
index $\alpha=2.3$ to describe the halo \citep{b36}, with the number
of halo to thick disk stars fixed within the Galaxy. In Figure
\ref{trend} we show the number distribution generated by this model
for the WGN fields and selection function.

In the top panel of Figure \ref{trend} $h_Z$ is kept fixed at 0.9 kpc
and $h_R$ varied from 2 to 8 kpc, while in the middle panel $h_R$ is
fixed at 4.0 kpc while $h_z$ is varied from 0.7 to 1.3 kpc.  From the
plots, we can see that the number counts at lower Galactic latitudes
are the most sensitive to thick disk structural parameters - as the
thick disk scale length is increased (or, with less effect, the scale
height is decreased), the number of stars near RA$\sim$135$^\circ$ and
RA$\sim$315$^\circ$ increases.  The data are mostly sensitive to the
thick disk radial scale length and a sufficient range of fields has
been targeted to provide useful constraints on those structural
parameters. This analysis thus illustrates the utility of the range of
fields selected for observation by WGN as a test of thick disk
structural parameters.

We define two figures of merit, Ratio1, being the ratio of the number
of stars per square degree selected by the WGN criteria in the field
at RA=313$^\circ$ to the corresponding number for the field at
RA=249$^\circ$; and Ratio2, the corresponding number comparing the
star counts in the field at RA=133$^\circ$ to those in the field at
RA=249$^\circ$. These figures of merit quantify how the change in the
thick disk structural parameters $h_R,h_Z$ affect the shape of the
star counts distribution as a function line of sight. The star count
distributions produced by the  Besan\c{c}on model has Ratio1=0.66 and
Ratio2=0.20 while the observed distribution from SDSS data has
Ratio1=1.69 and Ratio2=0.37. This quantifies the discrepancy between
data and model evident in Figure \ref{numbercounts}.

\begin{figure} 
\begin{center} 
\includegraphics[width=1.0\linewidth]{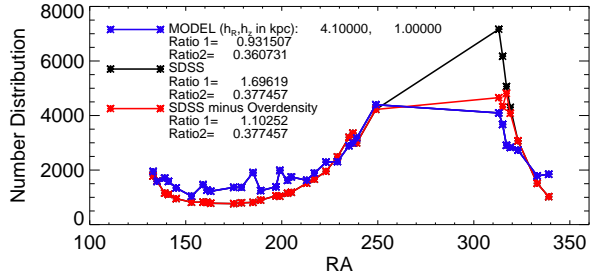} 
\caption{Comparison of SDSS photometric star count data in the WGN
  fields and the simple star count model described in the text. In the
  region RA$\le 250^{\circ}$ the star count data (black) and the simple
  model (blue) are in tolerable agreement. This indicates a
  photometric thick disk model with $h_R,h_Z$ = 4.1,1.0  kpc is an
  adequate description of the Galaxy. For RA$\ge 300^{\circ}$ the SDSS
  star count data (black) are significantly in excess of the model
  (blue), indicating an additional structure in the Galaxy. The red
  curve illustrates the SDSS count data after subtraction of a model
  for the excess counts described in the text and in Figure
  \ref{OVER_FIT}. This ``thick disk plus over-density'' model is an
  adequate description of the underlying ``normal'' thick disk
  population.}
\label{trial} 
\end{center} 
\end{figure}

By experiment we see that $(h_R,h_Z) = $ 2.1,1.0 kpc reproduces the
 Besan\c{c}on number distribution well (shown in bottom panel of Figure
\ref{trend}). That is, the  Besan\c{c}on model description of the thick
disk, for the observational parameters of relevance here, may be
described as a double exponential with $(h_R,h_Z) = $ 2.1,1.0 kpc. This is consistent with values quoted in \citet{b18} as used in the model for the thick disk density profile which is a modified exponential having scale length of 2.5$\pm$0.5 kpc, and scale height of 0.8$\pm$0.5 kpc. As
shown in Figure \ref{numbercounts}, however, that model substantially
and systematically under-predicts the observed SDSS star
counts. Increasing the radial thick disk scale length to the value
$h_R =$ 4.1 kpc, as shown in Figure \ref{trial}, makes the shape of
the number distribution consistent with SDSS data.This number is in
agreement with the value obtained by \citet{b9} who analyse M dwarfs
in the solar neighbourhood (at distances less than 2 kpc) from SDSS
data to estimate a scale length of 3.6 kpc for the thick
disk. While most previous works which estimate the
    scale length using star count analysis obtain scale length values
    between 3.5 and 4.5 kpc which is consistent with our analysis
    here, those works which have employed population separation based
    on a kinematic or chemical selection, such as \citet{TB2011} and
    \citet{Cheng2}, estimate a much lower scale length of around
    2kpc. Understanding this difference will no doubt significantly
    improve definition of the astrophysical nature of the thick disk.

While an increased $h_R$ reproduces the observed value of Ratio2,
there is still a large disagreement in Ratio1 (compare black line
(data) and blue line (model) in Figure \ref{trial}). Subtracting from
the data the contribution of the high velocity substructure determined
from Figure \ref{OVER_FIT} (red line), we see that the match improves
for Ratio1 too.

Thus, from this simple toy-model analysis we deduce a value of $\sim$
4.0 kpc for the thick disk scale length. We find that a smooth
double-exponential is, however, not a complete description of the data
and a substantial overdensity, compared to this model, is apparent in
the star counts, localised near RA$\sim$313$^\circ$. This overdensity
matches the fraction of stars contributed by the high-velocity
sub-structure identified in the previous section.

In the next section, we analyse and characterise the stars in this sub-structure. 

\subsection{Characterising the stars in the thick disk substructure}

\begin{figure} 
\begin{center} 
\includegraphics[width=1.0\linewidth]{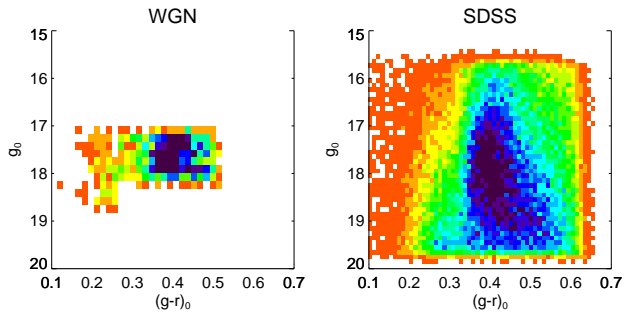} 
\caption{Left : Colour-magnitude diagrams from WGN data for two
  fields, covering the region $312^\circ<\mathrm{RA}<316^\circ$ and
  $-1^\circ<\mathrm{Dec}<1^\circ$. Binsize for this CMD is $\Delta
  g-r,\Delta g$= 0.015,0.15. Right : CMD from SDSS DR7 data for a
  larger region around the same fields covering
  $310^\circ<\mathrm{RA}<318^\circ$ and
  $-3^\circ<\mathrm{Dec}<3^\circ$. Binsize for this CMD is $\Delta
  g-r,\Delta g$= 0.01,0.1. The stars from the overdensity overlap with
  the thick disk turn off stars and are indistinguishable from thick
  disk. The colours denote low to high density from red to blue}
\label{OVER_CMD} 
\end{center} 
\end{figure}

In Figure \ref{OVER_FIT} we have already shown that the most
significant contribution of stars which have high velocity compared to
the model fit is restricted to just two fields in WGN (spanning around
4$^\circ$ in RA), beyond which the uniform thick disk model well
reproduces the line of sight velocity distributions.

\begin{figure} 
\begin{center} 
\includegraphics[width=1.0\linewidth]{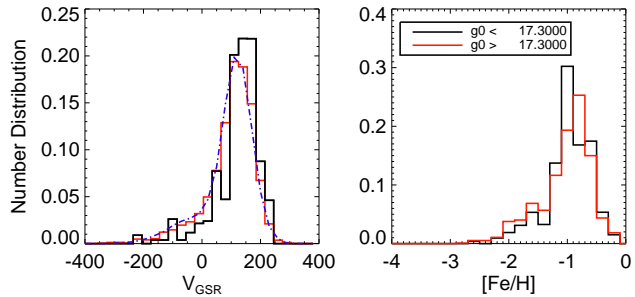} 
\caption {Unweighted line of sight velocity histograms are shown for
all stars in three fields from WGN covering
$312^\circ<\mathrm{RA}<318^\circ$. The sample is divided into brighter and
fainter than $g_0=17.3$ The left panel shows the line of sight
velocities (histograms) with the blue dot-dash curve showing the
expected model distribution.  The right panel shows the metallicity
distributions. The excess stars are most apparent with velocities
somewhat higher than that of the thick disk, with metallicity near
[Fe/H]=-1, and at brighter magnitudes.}
\label{OVER_SPLIT} 
\end{center} 
\end{figure}

Figure \ref{OVER_CMD} shows the colour-magnitude diagram of these two
fields in the region R313 where the contribution of the high velocity
stars is most significant. We see that both the WGN and SDSS DR7 CMD
look thick disk-like with the high-velocity stars overlapping with
thick disk stars. No special feature in the CMD or extra turn-off is
apparent. The substructure has a spread in apparent magnitudes as
indicated by Figure \ref{OVER_SPLIT}, but is preferentially seen at
brighter magnitudes, being most prominent at magnitudes brighter than
 $g \sim 17.3$. The null hypothesis that the two
    curves (black and red) in Figure \ref{OVER_SPLIT} are drawn from the same distribution is rejected at the 5
    percent level based on the Kolmogorov Smirnov Test. The same
    comparison was done for RA$\sim$133 and the null hypothesis was
    accepted.

Figure \ref{OVER_SPLIT} also reveals that the
substructure has metallicity somewhat below the thick disk, with mean
metallicity around -1.0 dex (black line).

\begin{figure*} 
\begin{minipage}{180mm} 
\begin{center} 
\includegraphics[width=0.7\linewidth]{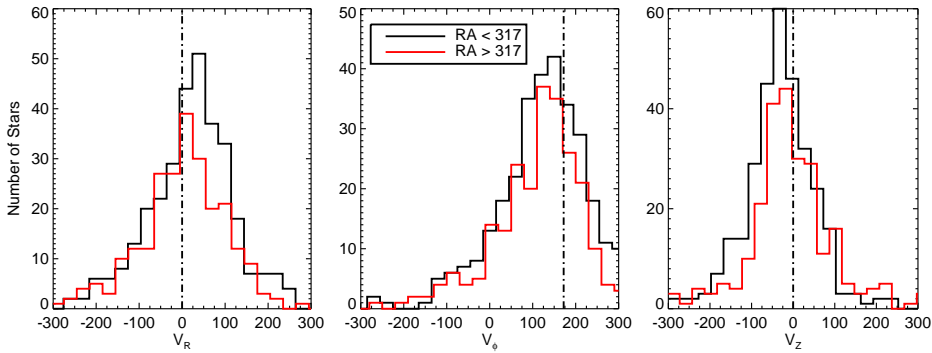} 
\caption{Galactocentric velocity distributions for the region with the
  star-count excess, R313 (black histogram), is compared to
  distributions from fields with $\mathrm{RA}>317^\circ$ which do not
  show a star count excess (red). The $V_\phi$ and $V_Z$ distributions
  are comparable, indicating a thick-disk-like velocity pattern, while
  the mean $V_R$ of the fields in R313 is offset towards positive
  values.}
\label{OVER_PM} 
\end{center} 
\end{minipage} 
\end{figure*}

Is it reasonable for us to assume the excess stars are in thick
disk-like orbits?  We looked at the $V_R,V_\phi,V_Z$ distribution of
stars from this line of sight by cross matching the WGN data with
proper motions from Stripe 82 based on position, being aware from our
analysis above (Figure \ref{erroranal}) that such comparisons are
indicative rather than conclusive. We used new high-precision values of SDSS star proper
motions re-calibrated by \citet{PMDatainprep} using background
galaxies. 613 matches were obtained.  Stars with proper motion errors
greater than 30mas/yr were removed. Figure \ref{OVER_PM} shows the
velocity distributions for the region of interest
($312^\circ<\mathrm{RA}<317^\circ$) compared to distributions from
nearby fields with no excess ($\mathrm{RA}>317^\circ$). The $V_\phi$
and $V_Z$ distributions of the two sets of fields are
indistinguishable. This is consistent with the excess stars being in
thick disk-like orbits. Whereas, the mean $V_R$ of the fields in R313
is offset towards positive values.

\subsection{Complementary and pre-discovery studies}

\begin{figure*} 
\begin{minipage}{180mm} 
\begin{center} 
\includegraphics[width=1.0\linewidth]{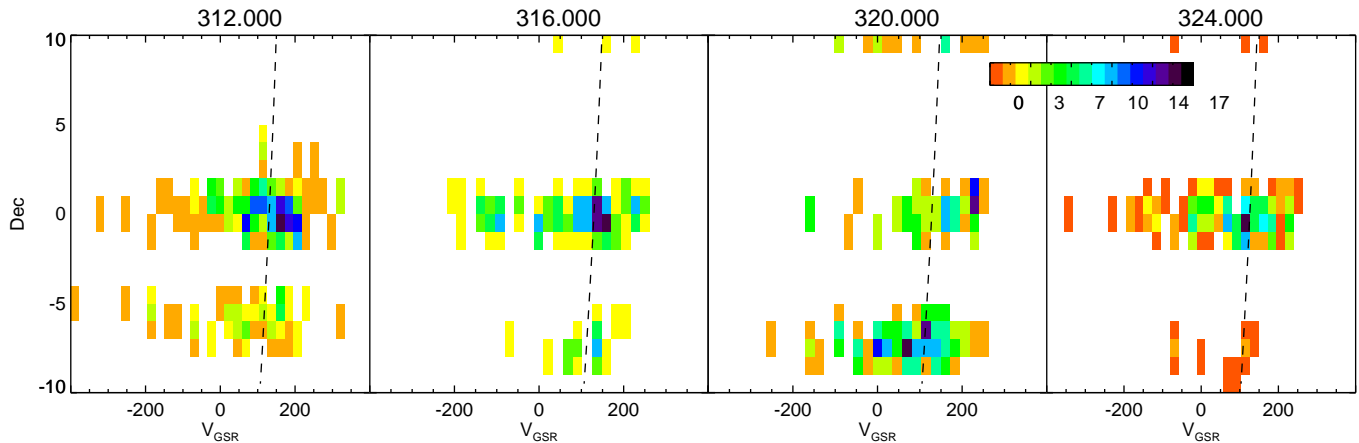} 
\caption{Radial velocities from SDSS DR7 spectra for a region
  $312^\circ<\mathrm{RA}<326^\circ$ and
  $-10^\circ<\mathrm{Dec}<10^\circ$, applying the WGN selection
  function. Each panel is a 4$^\circ$ block in RA, the mid value of RA
  being shown at the top of each panel. The colours represent low to
  high density from red to blue. The dashed line shows the position of
  thick disk rotation (i.e. $V_\phi$=172 km/s) at a distance of 2 kpc
  for each direction. We can see that there is an excess of stars at
  high V$_{GSR}$, that is here, close to zero V$_{helio}$ velocities
  (immediately to the right of the dashed line) in the first two
  panels, consistent with the excess stars seen in the WGN
  data.}
\label{OVER_DR7} 
\end{center} 
\end{minipage} 
\end{figure*}

An independent source of kinematics in the region of interest comes
from other SDSS studies, the SDSS spectra themselves. The SDSS DR7 spectra, selected applying the WGN selection function,
hint at a cold high velocity structure which is spatially and
kinematically consistent with R313. Figure \ref{OVER_DR7} shows line
of sight velocities from SDSS DR7 for a region
$312^\circ<\mathrm{RA}<326^\circ$ and
$-10^\circ<\mathrm{Dec}<10^\circ$. Each panel is a 4$^\circ$ block in
RA, the mid value of RA being shown at the top of each panel. The
colours represent low to high density from red to blue while the
dashed line shows the position of thick disk rotation (i.e
$V_\phi$=172 km/s) at a distance of 2 kpc for each direction and
serves as a reference. If velocities are consistent with being thick
disk, we would expect the peak in density in each bin of declination
to coincide with the dashed line. We can see that this is the case in
most bins, apart from an excess of stars at relatively high V$_{GSR}$
velocities (immediately to the right of the dashed line) in the first
two panels at Dec$\sim$ 0$^\circ$. This is consistent with the high
velocity stars seen in the WGN data. For clarity, recall that these
stars have observed radial velocity V$_{helio}$ near zero (cf fig
\ref{VRvsRA}). Any selection effects in SDSS
    spectroscopy are unlikely to vary enough
    from field to field, within the tight cuts of the WGN selection
    function, to change this result.

The overdensity observed by us is in a direction of the sky that is
extremely busy, with previously identified sub-structures such as the
Hercules Thick Disk Overdensity at 1-2 kpc, Hercules Aquila Cloud at
10-20kpc and more recent sub-structure observed in RR Lyrae stars in
Stripe 82 at 10-25kpc.
  
The Hercules Thick Disk Overdensity, first identified and studied in
detail by \citet{b15}, and further investigated in subsequent papers
(\citet{b16}, \cite{b16a}, \cite{b12}) thereafter, is an excess of
faint blue stars in the region $\ell=20^\circ-55^\circ$ and $|b|=20^\circ-45^\circ$, thus
overlapping with R313. These stars were estimated to have distances
between 1-2 kpc from the sun, consistent with distances estimated by our method for our
stars. \citet{b12} additionally also provide velocity measurements for
their fields. They found that stars in Quadrant I (where the excess
stars were found) had a $V_{LSR}$ of -28.6 $\pm$ 1.6 km/s above the
plane and a $V_{LSR}$ of -35.6 $\pm$ 2.0 km/s below the plane.  They
also state that going beyond 4 kpc along the line of sight, where the
overdensity coincides with the density contours of the Galactic bar, the
rotational rates become larger than standard thick disk
rotation. These kinematic parameters are in agreement with those we
find here, though our sample here is very much more spatially
concentrated than is that of \citet{b12}.

The Hercules Aquila Cloud \citep{VB07} is thought to be a large
structure further out in this same direction, centred on
$l\sim40^{\circ}$ and extending both above and below the plane up to
$b\sim50^{\circ}$. The cloud was estimated to be more metal rich than
-2.2 dex and have distances in the range of 10-20kpc. Their paper
quotes a corresponding kinematic signature of $V_{GSR}=180$km/s
associated with this cloud which matches the kinematic signal we find
in the WGN data.

We also looked at the density of RR Lyrae stars in Stripe 82
\citep{b82}, and towards the Hercules-Aquila density enhancement. In
Figure 11 of \cite{b82}, the overdensity marked B is centred on
similar directions as our overdensity, but extends between 5 and
25kpc. Cross-matching these stars with Sloan spectroscopy resulted in
a handful of matches. While these stars showed a peak in [Fe/H] near
-1.5dex, we did not find a coherent velocity for stars in this
structure.

Given there are at least two known overdensities in this direction at
distances beyond 10kpc, we investigated the possibility of the stars
in our overdensity in fact being at distances beyond 10kpc, but being
assigned grossly underestimated distances by our by our analysis techniques. Looking at the predictions from the Besan\c{c}on model, we find that few
sub-giants at large distances (further than 10kpc) pass through the
WGN selection function and such stars fall in the very red, faint end
of the WGN selection box. Our distance determination will indeed
grossly underestimate the distances to these stars, putting them at
2-4 kpc. But this would require that all of the stars in the
overdensity be clearly concentrated in colour-magnitude space at the
faint-red end. As seen from Figure \ref{OVER_CMD} this is not the
case. The overdensity stars instead look very thick-disk like.  

\section{Conclusions} 
 
We analyse the WGN medium resolution AAOmega spectroscopic survey of
some 10,000 stars, selected  using SDSS photometry to be main sequence turn-off stars in the thick
disk-halo interface. The goal has been to investigate large-scale and
small-scale kinematic, chemical and spatial properties of these two
old stellar populations, and additionally also investigate the region
of transition/overlap between these two populations.

Preparatory to the analysis, we developed a weighting method in
colour-magnitude space for completeness correction. Applying this
correction, we obtained a completeness-corrected representation of the
parent stellar population selected by the WGN selection function, thus
preparing the data for analysis. Robust distance estimates were
obtained by isochrone fitting, using either or both of SDSS $gri$ and
UKIDSS JHK photometry, for the stars, providing distances with a
systematic uncertainty of less than 10\%.
 
We demonstrated how it is convenient to model the data in observable
space directly. Propagated errors in derived parameters, especially
those in $V_R, V_\phi, V_Z$ due to proper motions, can be dominant,
and hence deconvolution can degrade the most accurate data, here
radial velocities and [Fe/H] values, from any analysis. Hence, we
analysed the kinematics of the thick disk using only observable line
of sight velocities, adopting the technique from \citet{b20} and
limiting the use of derived parameters to that of distance alone. In
this study, we obtain a global description of the thick disk
kinematics to demonstrate the scope of the data and the efficiency of
the technique. The best fit thick disk kinematic parameters are
determined to be $V_{\phi}$=172 km/s and $\sigma(\phi,r,z)$ =
(49,51,40) km/s which agrees reasonably well with values from previous
studies.
 
By comparing this global model description with the data, we identify
a region near RA$\sim$313$^{\circ}$ which has higher line of sight
V$_{GSR}$ velocities than expected for the thick disk in this
region. The same region is found also to have a local overdensity of stars
in comparison to that expected for thick disk density models. The
stars in the overdensity have a turn-off CMD similar to that of the
thick disk and metallicities $\sim -1$~dex, similar to, but slightly more metal poor
than, the thick disk \citep{b4}.  The stars in the excess are thus as old as are those of the
thick disk. The overdensity decreases rapidly over a 4$^\circ$
increase in RA from 312$^\circ$. Its further extent is poorly
determined as yet.

This overdensity spatially may be associated in some way with the
Hercules thick disk overdensity. Line of sight velocity measurements
from \citet{b12} match with the overdensity found in our WGN data, but
they infer lower rotational velocities than thick disk for the stars
in the overdensity while given our distance range we infer the
opposite. We should have also expected the velocity signature to
continue till at least RA of 320$^\circ$ if associated with the
Hercules thick disk overdensity which is a large structure.
This line of sight is towards the Hercules-Aquila
    halo cloud and a small spatial-scale overdensity in the RR Lyrae
    distribution. Obtaining reliable velocity estimates for these
    structures would be illuminating.

We have shown that the method of modelling in observable space is very
powerful not only to obtain a global description of the Galactic
stellar populations, but also to identify and pick out
substructure. We will extend this method in future work, to include
scientifically important questions of spatial gradients in structure,
kinematics and metallicity in the thick disk.

\section*{Acknowledgments}
AJ thanks The Boustany Foundation and Isaac Newton Trust for the award of scholarships. VB acknowledges financial support from the Royal Society. RFGW acknowledges support from National Science Foundation grants AST-0908326 and CDI-1124403, and thanks the Aspen Center for
Physics, supported by the National Science Foundation grant PHY-1066293, for hospitality.
.


\label{lastpage} 
\end{document}